\documentclass[onecolumn,preprintnumbers,amsmath,amssymb]{revtex4}

\usepackage{graphicx,array,color,epstopdf}

\usepackage{mathtools}
\usepackage[caption=false]{subfig}

\begin{document}

\title{Model reduction for stochastic chemical systems with abundant species}

\author{Stephen Smith}
\affiliation{School of Biological sciences, University of Edinburgh, Mayfield Road Edinburgh EH93JR Scotland, UK}

\author{Claudia Cianci}
\affiliation{School of Biological sciences, University of Edinburgh, Mayfield Road Edinburgh EH93JR Scotland, UK}

\author{Ramon Grima}
\affiliation{School of Biological sciences, University of Edinburgh, Mayfield Road Edinburgh EH93JR Scotland, UK}

 \begin{abstract}
Biochemical processes typically involve many chemical species, some in abundance and some in low molecule numbers. Here we first identify the rate constant limits under which the concentrations of a given set of species will tend to infinity (the abundant species) while the concentrations of all other species remains constant (the non-abundant species). Subsequently we prove that in this limit, the fluctuations in the molecule numbers of non-abundant species are accurately described by a hybrid stochastic description consisting of a chemical master equation coupled to deterministic rate equations. This is a reduced description when compared to the conventional chemical master equation which describes the fluctuations in both abundant and non-abundant species. {We show that the reduced master equation can be solved exactly for a number of biochemical networks involving gene expression and enzyme catalysis, whose conventional chemical master equation description is analytically impenetrable}. We use the linear noise approximation to obtain approximate expressions for the difference between the variance of fluctuations in the non-abundant species as predicted by the hybrid approach and by the conventional chemical master equation. Furthermore we show that surprisingly, irrespective of any separation in the mean molecule numbers of various species, the conventional and hybrid master equations exactly agree for a class of chemical systems.
\end{abstract}

\maketitle

 \section{Introduction}
The chemical master equation (CME) is the accepted stochastic description of chemical reaction systems \cite{VKbook}. Since intrinsic noise roughly scales as the inverse square root of the mean number of molecules \cite{VKbook}, it follows that the CME provides a more accurate description than deterministic rate equations (REs), when species are in low concentrations. However, exact solution of the CME has proved to be impossible for all but the simplest systems (see for example \cite{grima2012steady,shahrezaei2008analytical,smith2015general,jahnke2007solving}), and Monte Carlo simulations using the stochastic simulation algorithm (SSA) \cite{Gillespie1977} are also time-consuming in many cases of interest. One way to bypass these issues is to use a hybrid model which treats some parts of the system using the SSA and the rest using a simulation method which is computationally more efficient. {A common example of such hybrid modelling utilises time scale separation whereby some reactions occur on a fast timescale and are modelled using continuous approaches such as REs or chemical Langevin equations, while the rest of reactions occur on slow timescales and are modelled using the standard SSA \cite{haseltine2002,takahashi2004,salis2005}. Other methods which enable a considerable improvement over the SSA when time scale separation is present are: the nested SSA \cite{weinan2005nested}, a coarse-grained equation-free approach \cite{erban2006gene}, the constrained multiscale algorithm \cite{cotter2011constrained,cotter2015constrained}, an approach based on finite-state projection \cite{peles2006,munsky2006finite}, the slow-scale SSA \cite{Cao2005} and the slow-scale linear-noise approximation {\cite{Thomas2012a,thomas2012b}}. 
}

In this paper we consider a different type of hybrid model, one which uses a separation in the abundance of species (abundance separation) rather than timescale separation. In particular, we no longer split reactions into fast and slow, but rather categorise species based on how abundant they are. These methods utilise a continuous approach to model the abundant species and a discrete approach to model the less abundant species. While less popular than timescale separation, some hybrid algorithms have been developed to take advantage of this idea (see for example \cite{hellander2007,alfonsi2005,Jahnke2011,Jahnke2012}). Stochastic simulations verify that these hybrid models can capture important features of the fully stochastic model. In particular, the model by Hellander and Lotstedt \cite{hellander2007} has been shown by Jahnke \cite{Jahnke2011} to be exact for monomolecular systems, i.e., the marginal distributions of non-abundant species in the hybrid model are exactly the same as the same obtained from the full stochastic model. More sophisticated (and computationally expensive) approaches have been postulated \cite{Jahnke2011,Jahnke2012} to deal with systems which are not well described by the Hellander and Lotstedt hybrid model.

{The advantages of methods using abundance separation over timescale separation are that: (i) the timescales of reactions are often unknown while the abundances are readily measurable; (ii) there is evidence suggesting that abundance separation is at least as significant, if not more, than timescale separation for biochemical networks inside cells. For example it has been shown that the mean number of proteins per E. Coli cell is roughly a thousand times that of the mean number of mRNA molecules per cell \cite{taniguchi2010quantifying} -- in contrast, the ratio of protein to mRNA lifetime in E. Coli is expected to be above 1 but in the single digits \cite{shahrezaei2008analytical}. For mammalian cells, the same has been found; the median number of protein per cell is roughly 3000 times that of median number of mRNA per cell -- in contrast, the ratio of median protein to median mRNA lifetime is about 5 \cite{Schwan2011}. Clearly in these cases, abundance separation is significant while timescale separation is weak, and thus a method which takes advantage of the former appears to be ideal as a means to infer information about the stochastic dynamics of mRNA and of other proteins present in low copy numbers.}  

In this paper we postulate a novel simple hybrid model based on abundance separation consisting of a CME for the non-abundant species coupled to REs for all species. Subsequently we identify the rate constant limits under which the concentrations of a given set of species will tend to infinity (the abundant species) while the concentrations of all other species remain constant (the non-abundant species). This limit we shall refer to as the abundance or abundant limit. We show that in this limit, the marginal distributions of the non-abundant species given by the hybrid model converge to the same distributions given by the CME of the full system. {This fact is particularly useful when the hybrid model can be solved analytically, which is the case in several examples that we study.} We illustrate the accuracy of our hybrid model by comparing the exact stochastic simulations of its reduced CME with exact stochastic simulations of the full CME. We also show that there are several chemical systems for which our hybrid model is exact even without abundance separation. In Section 4 we offer an error analysis which provides an easy means to estimate the error incurred by the use of the hybrid model when the ratios of abundant to non-abundant species concentrations are finitely large. We conclude with a Summary and Discussion in Section 5.

\section{\label{sec:level1}A reduced chemical master equation}

In this section, we first propose a reduced CME which constitutes our hybrid model, and subsequently rigorously prove that it converges to the CME of the full system (that describing all species) in the abundance limit.

\subsection{A heuristic reduction of the CME}

The CME for a system of $N$ chemical species which interact through $R$ reactions has the form:
\begin{align}\label{ME}
\frac{d}{dt}P(\bold{n},t)=\Omega \sum_{j=1}^R \left( \prod_{i=1}^N E_i^{-S_{ij}} -1 \right) \hat{f}_j(\Omega,\bold{n})P(\bold{n},t),
\end{align}
where $\Omega$ is the volume in which the reactions occur, $E_i^{x}$ is the step operator which replaces $n_i$ with $n_i+x$, the entries of the state vector $\bold{n}=(n_1,...,n_N)$ are the number of molecules of each species, $P(\bold{n},t)$ is the probability of the system being in state $\bold{n}$ at time $t$, and the stoichiometric matrix elements $S_{ij}$ are given by the net change in the number of molecules of species $i$ when the $j$th reaction occurs. The probability that a reaction $j$ occurs in an time interval $[t,t+dt)$ is given by $\Omega \hat{f}(\bold{n},\Omega)dt$, where $\hat{f}_j$ is a function of elements of the state vector and reaction rates. The REs are defined by:
\begin{equation}\label{redef}
\frac{d{\phi}_i}{dt}=\sum_{j=1}^R S_{ij}f_j(\vec{\phi}),
\end{equation}
where $\vec{\phi}=(\phi_1,...,\phi_N )$ are the concentrations of the $N$ chemical species, and $f_j(\vec{\phi})=\text{lim}_{\Omega \rightarrow \infty}\hat{f}_j(\Omega,\Omega \vec{\phi} )$. 

We wish to reduce the number of species in this CME from $N$ to $M$ with $M<N$. Without loss of generality, we will keep species 1 to $M$, and remove species $M+1$ to $N$, which we consider to be ``abundant". We will do this by first summing the CME over $n_{M+1},~...,~n_N$ to leave us with an equation for the time evolution of the exact marginal distribution $P^{*}(\bold{n'},t)$, where $\bold{n'}=(n_1,...,n_M)$, then subsequently we use a mean-field assumption to obtain a time evolution equation for the approximate marginal distribution $\tilde{P}(\bold{n'},t)$. 

We will be considering the different possible forms that $\hat{f}_j$ can take assuming elementary reactions, specifically up to bimolecular reactions (reactions involving more than three molecules are rare in a biological setting). We will first investigate what happens to the CME if we sum over, say, $n_h$, using the notation that $\bold{n}_{-h}$ is the state vector $\bold{n}$ without the $h^\text{th}$ entry, in other words, $\bold{n}_{-h}=\left(n_1,...,n_{h-1},n_{h+1},...,n_N\right)$. In what follows, we will use $\bold{X}=\left(X_1,...,X_N\right)$ to refer to the vector of chemical species, and $\bold{Y}=\left(Y_1,...,Y_N\right)$ to refer to the vector of  random variables which give the number of molecules of each species. The state vector $\bold{n}=\left(n_1,...,n_N\right)$ therefore refers to a particular realisation of the random vector $\bold{Y}$.

If reaction $j$ does not feature $X_h$ amongst its reactants, then $\hat{f}_j$ has no $n_h$ dependence and the corresponding term in the CME remains unchanged.

If reaction $j$ is a unimolecular reaction of the type $X_h\rightarrow ...$ then $\hat{f}_j(\Omega,\bold{n})=k_jn_h\Omega^{-1}$, and we will have:
\begin{align}
\sum_{n_h=0}^{\infty}k_jn_h\Omega^{-1}P(\bold{n},t)=k_j\Omega^{-1} \langle Y_h \vert Y_1=n_1,Y_2=n_2,...\rangle  P(\bold{n}_{-h},t).
\end{align}

If reaction $j$ is a bimolecular reaction of the type $X_h+X_h \rightarrow ...$ then $\hat{f}_j(\Omega,\bold{n})=k_jn_h(n_h-1)\Omega^{-2}$ and we will have:
\begin{align}
\sum_{n_h=0}^{\infty}k_jn_h(n_h-1)\Omega^{-2}P(\bold{n},t)=k_j\Omega^{-2} \langle Y_h (Y_h-1)\vert Y_1=n_1,Y_2=n_2,...\rangle  P(\bold{n}_{-h},t).
\end{align}

Finally, if reaction $j$ is a bimolecular reaction of the type $X_h+X_i\rightarrow ...$ then $\hat{f}_j(\Omega,\bold{n})=k_jn_hn_i\Omega^{-2}$ and we will have:
\begin{align}
\sum_{n_h=0}^{\infty}k_jn_hn_i\Omega^{-2}P(\bold{n},t)=k_j\Omega^{-2}n_i \langle Y_h\vert Y_1=n_1,Y_2=n_2,...\rangle  P(\bold{n}_{-h},t).
\end{align}

These results follow from the definition of conditional expectation:
\begin{equation}
\langle f(X) \vert  Y = y \rangle = \sum_{x} f(x) \frac{P(X = x, Y = y)}{  P(Y = y)}. 
\end{equation}

Given the above results, we can now see what will happen to each $\hat{f}_j$ if we sum over all $h=M+1,...,N$. The result of this summation leads to new propensities involving conditional expectations which we call $f_j^\star$.  It then follows that the exact marginal CME is given by:
\begin{align}\label{exmacme}
\frac{d}{dt}P^\star (\bold{n'},t)=\Omega \sum_{j=1}^R \left( \prod_{i=1}^M E_i^{-S_{ij}} -1 \right) f^\star_j(\Omega,\bold{n})P^\star (\bold{n'},t),
\end{align}
where $\bold{n'}=(n_1,...,n_M)$. In theory, we have simplified the CME while keeping it exact, but we should be careful because the dependence of the conditional expectations on $\bold{n}'$ is currently unknown.

\begin{table}
\centering
    \begin{tabular}{ | l | l |}
    \hline
      Propensity $\hat{f}_j(\Omega,\bold{n})$& Reduced propensity $\tilde{f_j}(\Omega,\bold{n'},\vec{\phi}(t))$\\ \hline
      $k_j$& $k_j$\\ \hline
    $k_jn_i\Omega^{-1} $ & $ k_jn_i\Omega^{-1}$\\ \hline
$k_j n_i n_r \Omega^{-2}$ & $k_j n_i n_r \Omega^{-2}$\\ \hline
  $k_jn_i(n_i-1)\Omega^{-2}$&$ k_jn_i(n_i-1)\Omega^{-2}$\\ \hline
   $k_jn_i n_h\Omega^{-2}$ &$k_jn_i\Omega^{-1}\phi_h(t)$ \\ \hline
   
$k_jn_h\Omega^{-1}$ &$  k_j\phi_h(t)$ \\\hline
$k_jn_h(n_h-1)\Omega^{-2}$  & $  k_j\phi_h^2(t)$ \\\hline
$k_jn_hn_v\Omega^{-2}$  & $k_j\phi_h(t)\phi_v(t)  $\\\hline 
    \end{tabular}\\
\caption{The recipe for converting standard propensities $\hat{f}_j$ to reduced propensities $\tilde{f}_j$. The subscripts $i,r\leq M$ refer to non-abundant species, while $h,v>M$ refer to abundant species.}
\end{table}

We proceed by making the heuristic mean-field assumption that these conditional expectations can be replaced by deterministic concentrations $\phi_i$ as defined earlier in Eq. \eqref{redef}, for example,
\begin{align}
\langle Y_h \vert Y_{1}=n_1,...,Y_M=n_M \rangle \approx \Omega \phi_h,
\end{align}
where we have approximated away all conditional dependence. We can correspondingly update the exact effective propensities $f^\star_j$ with the approximate effective propensities $\tilde{f}_j$. A general recipe for converting $\hat{f}_j$ to $\tilde{f}_j$ is given in Table 1.

The approximate marginal CME is now:
\begin{equation}\label{apmarcme}
\frac{d}{dt}\tilde{P} (\bold{n'},t)=\Omega \sum_{j=1}^R \left( \prod_{i=1}^M E_i^{-S_{ij}} -1 \right) \tilde{f}_j(\Omega,\bold{n'},\vec{\phi}(t))\tilde{P} (\bold{n'},t).
\end{equation}

In the rest of the paper, we will refer to Eq. (\ref{ME}) as the full CME and Eq. (\ref{apmarcme}) as the reduced CME. 

An alternative way of summarising our reduction method, is that it consists of approximating a general chemical system:
\begin{align}
s_{1j}X_1+...+s_{Nj}X_N \xrightarrow{k_j} r_{1j}X_1+...+r_{Nj}X_N, \quad j = 1,..., R
\end{align}
by the reduced chemical system:
\begin{align}
\label{redchemsysN}
s_{1j}X_1+...+s_{Mj}X_M \xrightarrow{k_j\phi_{M+1}^{s_{M+1,j}}...\phi_{N}^{s_{N,j}}}r_{1j}X_1+...+r_{Mj}X_M, \quad j = 1,..., R,
\end{align}
when species $X_{M+1},...,X_N$ are abundant.

\subsection{The Abundant Limit}\label{AbunLim}

We wish to show that the approximate marginal CME given in Eq. \eqref{apmarcme} is a good approximation to the exact marginal CME given in Eq. \eqref{exmacme} when species $X_{M+1},...,X_N$ are abundant. To do this, we will need to define an abundance limit. Precisely, we want to know which parameters we should tweak in order that some species concentrations should go to infinity, while others stay constant. We will assume, without loss of generality, that we want to take the abundant limit of species $X_N$. For systems with multiple abundant species, we can just repeat the below process for each one in turn. 

The convention we use for numbering reactions is given in Table 2. We will introduce the functions $a(i)$ and $b(i)$ so that we can say that the bimolecular reaction with rate $k_i$ has species $X_{a(i)}$ and $X_{b(i)}$ as its reactants, where $a(i) \ne b(i)$. {We will have $N$ input reactions with rate $k_i,~i=-1,...,-N$ which lead to the production of each species}, monomolecular reactions with rates $k_i,~i=1,...,N$, bimolecular reactions between different species with rates $k_i,~i=N+1,...,L$ (for some $L \in \mathbb{N}$), and  bimolecular reactions between the same species with rates $k_i,~i=L+1,...,L+N$. 

\begin{table}
\centering
    \begin{tabular}{ | l | l | l |}
    \hline
    Reaction index & Reaction type & Reaction rate\\ \hline
   {$i=-1,...,-N$ }& $\emptyset \rightarrow ...$ & $k_i$ \\ \hline
   $ i=1,...,N$ &$ X_i \rightarrow ...$&$ k_i\phi_i$ \\ \hline   
    $i=N+1,...,L$ &$ X_{a(i)}+X_{b(i)} \rightarrow ...$ &$ k_i \phi_{a(i)}\phi_{b(i)}$ \\ \hline
  $i=L+1,...,L+N$ &$ 2 X_{i-L} \rightarrow ... $& $k_{i}\phi_{i-L}^2$\\ \hline
    \end{tabular}
\caption{A convention for numbering reactions. {Reactions $-1,...,-N$ are input reactions}, reactions $1,...,N$ are unimolecular, reactions $N+1,...,L$ are bimolecular and between different species and reactions $L+1, ..., L+N$ are bimolecular and between the same species.}
\end{table}

Now the rate equation for the concentration of $X_r$ is:
\begin{align}
\label{eqreN}
\frac{d}{dt} {\phi}_r=\sum_{i=-N}^{-1}S_{ri}k_i+\sum_{i=1}^N S_{ri}k_i\phi_i+\sum_{i=N+1}^L k_i S_{ri}\phi_{a(i)} \phi_{b(i)}+\sum_{i=L+1}^{L+N} S_{ri}k_{i}\phi_{i-L}^2,
\end{align}
where $S_{ri}$ is the net change in the number of molecules of species $X_r$ when reaction $i$ occurs.

{An intuitive means to obtain an abundant species $X_N$ is to make the rate constants of the reactions which remove this species, to be very small, whilst the rest of the rate constants remain at their constant value}. In particular, if exactly one molecule of $X_N$ is a reactant, then we let $k_j \propto \frac{1}{x}$; if two molecules of $X_N$ are reactants, then we let $k_j \propto \frac{1}{x^2}$, where $x \rightarrow \infty$. This means that $k_{L+N} \propto \frac{1}{x^2}$, $k_N \propto \frac{1}{x}$, $k_j \propto\frac{1}{x}$ for $j$ such that $a(j)$ or $b(j)$ equal $N$ and $j=N+1,...,L$. In what follows we shall refer to these rate constant limits as the abundance or abundant limit. Plugging in the aforementioned rates constant scalings and the trial solution:
\begin{align} 
\label{ss_sol}
 &\phi_i = c_i, \quad i \ne N \notag \\ &\phi_N= c_N x, 
 \end{align}
in Eq. (\ref{eqreN}) where $c_i$ are constants independent of $x$, and considering steady-state by setting the time derivative to zero, one obtains a set of $N$ simultaneous equations in the $N$ constants $c_i (i = 1, ..., N)$. Importantly the coefficients of these simultaneous quadratic equations are independent of $x$ which implies that if these equations can be solved for $c_i$ then the solution is independent of $x$, as they should indeed be, given the form of the trial solution above. Thus it follows that provided the simultaneous equations can be solved, the steady-state solution of the REs is given by Eq. (\ref{ss_sol}). Note this implies that that in the abundance limit, the ratio of the abundant to the non-abundant species concentrations, $\phi_N/\phi_i$ ($i \ne N$), scales as $x$ where $x \rightarrow \infty$, whilst the concentration of the non-abundant species remains constant. 

Note also that we have here implicitly assumed that there is no chemical conservation law which involves an abundant species {(chemical conservation laws which involve only non-abundant species are however allowed)}. This is since this law necessitates a finite upper bound on the concentrations which is contrary to the manner in which we have here defined the abundant limit. 

For systems with multiple abundant species, the above recipe implies that $k_j \propto \frac{1}{x^q}$ where $q$ is the total number of reactant molecules of abundant species involved in reaction $j$. For example for the reaction $X_{h}+X_{v} \rightarrow ...$ where both species $X_h$ and $X_v$ are abundant, the rate constant of the reaction scales as $\frac{1}{x^{2}}$.

The limits here derived assume a steady-state rate equation description for all species. This derivation is here presented to simplify the presentation and since it is very intuitive. However as we show in the next section, the limits elucidated here, also constitute abundance limits for the time-dependent stochastic description. 

\subsection{Proof of $N$ Species Abundant Convergence}
We will now use the limits derived in section 2.2 to prove that the approximate marginal distribution $\tilde{P}(\bold{n'},t)$ governed by the heuristic marginal CME Eq. \eqref{apmarcme} converges to the exact marginal distribution $P^\star (\bold{n'},t)$ governed by the exact marginal CME Eq. \eqref{exmacme} in the abundance limit. 

\subsubsection{Taylor expansion of exact marginal distribution}
A full $N$ species CME with $R$ reactions has the form of Eq. \eqref{ME}. We will expand the solution $P(\bold{n},t)$ as a Taylor series in time about $t=0$.
 We assume deterministic initial conditions, $P(\bold{n},t=0)=\delta^{n^{0}_{1}}_{n_1}...\delta^{n^0_N}_{n_N}$, where $n_i^0$ denotes the initial value of $n_i$. We can write the Taylor expansion:
\begin{align}
P(\bold{n},t)=\sum_{k=0}^{\infty}P^{(k)}(\bold{n},0)\frac{t^{k}}{k!},
\end{align}
where $P^{(k)}$ is the $k^\text{th}$ time derivative of $P$. Since the full CME is a coupled set of first-order ordinary differential equations with constant coefficients, the Taylor series above is guaranteed to have an infinite radius of convergence by Fuchs's theorem \cite{BenderOrszag}. 

From this series we can compute the marginal distribution:
\begin{align}
P^{\star}(\bold{n'},t)&=\sum_{n_{M+1}=0}^{\infty}...\sum_{n_N=0}^\infty P(\bold{n},t) \nonumber\\
&=\sum_{n_{M+1}=0}^{\infty}...\sum_{n_N=0}^\infty\sum_{k=0}^{\infty}P^{(k)}(\bold{n},0)\frac{t^{k}}{k!}\nonumber\\
&=~~\sum_{k=0}^{\infty}P^{\star (k)}(\bold{n}',0)\frac{t^{k}}{k!}.
\end{align}
We already know $P(\bold{n},0)=\delta^{n^{0}_{1}}_{n_1}...\delta^{n^0_N}_{n_N}$, so our first problem is the second term of the expansion, which is the first time derivative. This is given by the CME Eq. \eqref{ME} itself: 
\begin{align}
\dot{P}(\bold{n},0)&=\Omega \sum_{j=1}^R \left( E_1^{-S_{1j}}...E_N^{-S_{Nj}}-1 \right)\hat{f}_j(\Omega,\bold{n})P(\bold{n},0)\nonumber\\
&=\Omega \sum_{j=1}^R \left( E_1^{-S_{1j}}...E_N^{-S_{Nj}}-1 \right)\hat{f}_j(\Omega, \bold{n})\delta^{n^{0}_{1}}_{n_1}...\delta^{n^0_N}_{n_N},
\end{align}
and thus we can compute the $k^\text{th}$ derivative,
\begin{align}
P^{(k)}(\bold{n},0)=\Omega^k\left( \sum_{j=1}^R \left[ \left( E_1^{-S_{1j}}...E_N^{-S_{Nj}}-1 \right)\hat{f}_j(\Omega,\bold{n})\right] \right)^k \delta^{n^{0}_{1}}_{n_1}...\delta^{n^0_N}_{n_N},
\end{align}
where we are careful to note that $P^{(k)}(\bold{n},0) \not\equiv \left( \dot{P}(\bold{n},0) \right)^k$, since the $E$ operators do not commute with the propensities $\hat{f}_j$. 
Now we can compute the marginal distribution $P^\star$, which is made much simpler by the presence of the Kronecker-deltas:
\begin{align}\label{exmartay}
&P^{\star (k)}(\bold{n'},0)=\sum_{n_{M+1}=0}^{\infty}...\sum_{n_N=0}^\infty P^{(k)}(\bold{n},0) \nonumber\\
&=\Omega^k\left( \sum_{j=1}^R \left[ \left( E_1^{-S_{1j}}...E_M^{-S_{Mj}} E_{M+1'}^{-S_{M+1,j}}...E_{N'}^{-S_{Nj}}-1 \right)\hat{f}_j(\Omega,\bold{n'},n_{M+1}^0,...,n_N^0)\right] \right)^k \delta^{n^{0}_{1}}_{n_1}...\delta^{n^0_M}_{n_M},
\end{align}
where $E_{i'}^x$ now acts on the initial conditions $n_i^0$ rather than the variable $n_i$.

\subsubsection{Taylor expansion of the approximate reduced distribution}

The approximate marginal distribution $\tilde{P}(\bold{n}',t)$ is defined by the reduced CME:
\begin{align}\label{eqnap}
\dot{\tilde{P}}(\bold{n'},t)=\Omega \sum_{j=1}^R \left( E_1^{-S_{1j}}...E_M^{-S_{Mj}}-1 \right)\tilde{f}_j(\Omega,\bold{n'},\vec{\phi}(t))\tilde{P}(\bold{n'},t).
\end{align}
Its Taylor expansion about $t=0$ is:
\[
\tilde{P}(\bold{n}',t)=\sum_{k=0}^\infty \tilde{P}^{(k)}(\bold{n}',0) \frac{t^k}{k!}.
\]

Fuchs's theorem guarantees that a first-order ordinary differential equation with time-dependent coefficients will have a radius of convergence at least as large as the minimum of the radius of convergence of the time-dependent parameters \cite{BenderOrszag}. The reduced CME is a set of coupled first-order equations with time-dependent coefficients determined by the solution of the REs. Hence if the REs admit a Taylor series solution with an infinite radius of convergence then the Taylor series of the reduced CME also does. 

We already know $\tilde{P}(\bold{n'},0)=\delta_{n_1}^{n_1^0}...\delta_{n_M}^{n_M^0}$, so our first problem is the second term of the expansion. Again, this is given by the approximate CME Eq. \eqref{eqnap},
\begin{align}
\dot{\tilde{P}}(\bold{n'},0)&=\Omega \sum_{j=1}^R (E_{1}^{-S_{1j}}...E_M^{-S_{Mj}}-1)\tilde{f}_j(\Omega,\bold{n'},\vec{\phi}(0))\tilde{P}(\bold{n'},0)\nonumber\\
&= \Omega \sum_{j=1}^R (E_{1}^{-S_{1j}}...E_M^{-S_{Mj}}-1)\tilde{f}_j(\Omega,\bold{n'},\vec{\phi}(0))\delta_{n_1}^{n_1^0}...\delta_{n_M}^{n_M^0},
\end{align}
where we note that the propensities $\tilde{f}_j$ will in general depend on $t$ as well as $\bold{n'}$. This will cause complications, for instance, the second derivative has the form:
\begin{align}
\tilde{P}^{(2)}(\bold{n'},0)&=\Omega \sum_{j_1=1}^R(E_{1}^{-S_{1j}}...E_M^{-S_{Mj}}-1)\left[ \dot{\tilde{f}}_j(\Omega,\bold{n'},\vec{\phi}(0))\tilde{P}(\bold{n'},0)+\tilde{f}_j(\Omega, \bold{n'},\vec{\phi}(0))\tilde{P}^{(1)}(\bold{n'},0) \right]\nonumber\\
&= \Omega \sum_{j=1}^R (E_{1}^{-S_{1j}}...E_M^{-S_{Mj}}-1)\dot{\tilde{f}}_{j} (\Omega,\bold{n'},\vec{\phi}(0))\delta_{n_1}^{n_1^0}...\delta_{n_M}^{n_M^0}+\\
&~~~~\Omega^2 \sum_{j_1=1}^\infty \sum_{j_2=1}^\infty \left[ (E_{1}^{-S_{1j_1}}...E_M^{-S_{Mj_1}}-1) \tilde{f}_{j_1}(\Omega,\bold{n'},\vec{\phi}(0)) (E_{1}^{-S_{1j_2}}...E_M^{-S_{Mj_2}}-1) \tilde{f}_{j_2}(\Omega,\bold{n'},\vec{\phi}(0))\delta_{n_1}^{n_1^0}...\delta_{n_M}^{n_M^0}\right].\nonumber
\end{align}
We now have an extra term in our sum which depends on the time derivative of the $\tilde{f}_j$, and if we take higher order Taylor coefficients, we get higher time derivatives of the $\tilde{f}_j$. We get, using the notation used in the previous section,
\begin{align}\label{apmartay}
\tilde{P}^{(k)}(\bold{n'},0) = &\Omega^k\left( \sum_{j=1}^R \left[ \left( E_1^{-S_{1j}}...E_M^{-S_{Mj}}-1 \right)\tilde{f}_j(\Omega,\bold{n'},\vec{\phi}(0))\right] \right)^k \delta^{n^{0}_{1}}_{n_1}...\delta^{n^0_M}_{n_M} \nonumber\\ 
&+ \text{ terms proportional to the time derivatives of the } \tilde{f}_j.
\end{align}

\subsubsection{Convergence of full and reduced Taylor series}
The absolute difference between the $k^\text{th}$ terms of the two Taylor CMEs is given by the difference between Eqs. \eqref{exmartay} and \eqref{apmartay}:
\begin{align}
\Omega^k &\Bigg[ \left( \sum_{j=1}^R \left[ \left( E_1^{-S_{1j}}...E_M^{-S_{Mj}}-1 \right)\tilde{f}_j(\Omega,\bold{n'},\vec{\phi}(0))\right] \right)^k\nonumber \\
& -\left( \sum_{j=1}^R \left[ \left( E_1^{-S_{1j}}...E_M^{-S_{Mj}} E_{M+1'}^{-S_{M+1,j}}...E_{N'}^{-S_{Nj}}-1 \right)\hat{f}_j(\Omega,\bold{n'},n_{M+1}^0,...,n_N^0)\right] \right)^k \Bigg]\delta^{n^{0}_{1}}_{n_1}...\delta^{n^0_M}_{n_M}\nonumber\\
&+ \text{ terms proportional to the time derivatives of the } \tilde{f}_j.
\end{align}
This will tend to zero in the abundant limit if we can prove two claims:
\\\\
(I) For any $\alpha_i \in \mathbb{Z}$, and $j \in {1,...,R}$, 
\begin{align}
\left[ \tilde{f}_{j}(\Omega,n_1+\alpha_1,...,n_M+\alpha_M,\vec{\phi}(0))-\hat{f}_j(\Omega,n_1+\alpha_1,...,n_N^0+\alpha_N) \right] \rightarrow 0,
\end{align} in the abundant limit.
\\\\
(II) The time derivatives of the $\tilde{f}_j$ tend to zero in the abundant limit.
 \\\\
To prove (I), we must recall the different possible forms that $\hat{f}_j$ and $\tilde{f}_j$ can take, which are given in Table 3. As before, we say that the indices $i,r \leq M$ correspond to non-abundant species, while the indices $h,v>M$ correspond to abundant species.
\begin{table}
    \begin{tabular}{ | l | l | l |}
    \hline
    Reaction type & $\tilde{f}_j $&$ \hat{f}_j$ \\ \hline
    $\emptyset \rightarrow ...$ &$ k_j$ & $k_j$ \\ \hline
    $X_i \rightarrow ...$ & $k_j(n_i+\alpha_i)\Omega^{-1}$ &$ k_j(n_i+\alpha_i)\Omega^{-1}$ \\ \hline
    $ X_h \rightarrow ...$ &$ k_j \phi_h(0) $& $k_j(n_h^0+\alpha_h) \Omega^{-1}$\\ \hline
   $X_i+X_r\rightarrow ...$ &$ k_j (n_i+\alpha_i)(n_r+\alpha_r)\Omega^{-2}$ &$ k_j (n_i+\alpha_i)(n_r+\alpha_r)\Omega^{-2}$ \\ \hline
    $X_i+X_i\rightarrow ...$ &$ k_j (n_i+\alpha_i)(n_i+\alpha_i-1)\Omega^{-2}$ &$ k_j (n_i+\alpha_i)(n_i+\alpha_i-1)\Omega^{-2}$ \\ \hline
    $X_h+X_v\rightarrow ...$ & $k_j\phi_h(0)\phi_v(0) $&$ k_j (n_h^0+\alpha_h)(n_v^0+\alpha_v) \Omega^{-2}$\\ \hline
    $X_h+X_h\rightarrow ...$ & $k_j\phi^2_h(0) $&$ k_j (n_h^0+\alpha_h)(n_h^0+\alpha_h-1) \Omega^{-2}$\\ \hline
    $X_i+X_h\rightarrow ...$ &$ k_j  (n_i+\alpha_i) \phi_h(0) \Omega^{-1}$ &$ k_j (n_i+\alpha_i) (n_h^0+\alpha_h) \Omega^{-2}$\\
    \hline
    \end{tabular}
\caption{The different forms that $\tilde{f}_j$ and $\hat{f}_j$ can take in our Taylor-expanded distributions. The $\alpha_i$ correspond to integer changes in species number introduced by the $E_i^x$ operators. Note that indices $i,r \leq M$ correspond to non-abundant species, while the indices $h,v>M$ correspond to abundant species.}
\end{table}

Convergence is trivial for reactions $\emptyset \rightarrow ...$, $X_i \rightarrow ...$, $X_i+X_r\rightarrow ...$, and $X_i+X_i\rightarrow ...$, since $\tilde{f}_j$ and $\hat{f}_j$ agree. For the other reactions, we have to decide how the initial conditions should scale in the abundant limit. We will suppose that the initial concentration for species $X_M+1,...,X_N$ should tend to infinity $O(x)$ (since they are the abundant species), while the initial concentration for species $1,...,M$ should stay constant. 

For the $ X_h \rightarrow ...$ reaction, we are interested in
\begin{align}
k_j \phi_h(0)-k_j (n_h^0+\alpha_h ) \Omega^{-1}.
\end{align}
However by definition, $\phi_h(0)=\frac{n_h^0}{\Omega}$, so the absolute error becomes $\frac{k_j \alpha_h}{\Omega}$ which tends to zero since $k_j \rightarrow 0$ in accordance with our abundant limit elucidated in section 2.2.
\\\\
For the $X_h+X_v\rightarrow ...$ reaction, we are interested in
\begin{align}
k_j\phi_h(0)\phi_v(0)-k_j (n_h+\alpha_h)(n_v+\alpha_v) \Omega^{-2},
\end{align}
which simplifies to
\begin{align}
-k_j \left(\frac{n_h^0\alpha_v}{\Omega^2}+\frac{n_v^0\alpha_h}{\Omega^2}+\frac{\alpha_h\alpha_v}{\Omega^2} \right),
\end{align}
and which again tends to zero, since in the abundant limit $k_j \propto \frac{1}{x^2}$ while $n_h^0$ and $n_v^0$ are proportional to $x$ and $x \rightarrow \infty$.
\\\\
For the $X_h+X_h\rightarrow ...$ reaction, we are interested in,
\begin{align}
k_j\phi_h(0)^2-k_j (n_h^0+\alpha_h)(n_h^0+\alpha_h-1) \Omega^{-2},
\end{align}
which simplifies to
\begin{align}
-k_j\left(\frac{2n_h^0\alpha_h}{\Omega^2}-\frac{n_h^0}{\Omega^2}+\frac{\alpha_h^2-\alpha_h}{\Omega^2} \right),
\end{align}
and which again tends to zero since $k_j \propto \frac{1}{x^2}$ while $n_h^0 \propto x \rightarrow \infty$.
\\\\
Finally, for the $X_i+X_h\rightarrow ...$ reaction we consider
\begin{align}
k_j  (n_i+\alpha_i) \phi_h(0) \Omega^{-1} -k_j (n_i+\alpha_i) (n_h^0+\alpha_h) \Omega^{-2},
\end{align}
which we write as
\begin{align}
-\alpha_h k_j  (n_i+\alpha_i) \Omega^{-2},
\end{align}
and which also tends to zero since $k_j \rightarrow 0$ in the abundance limit. We have therefore proved claim (I).
\\\\
To prove claim (II), we will use the convention for reactions introduced in Table 2. We will say that $k_0$ corresponds to the rate of a null reaction, $k_1,...,k_N$ correspond to the rates of monomolecular reactions, $k_{N+1},...,k_{L}$ correspond to the rates of bimolecular reactions involving distinct species with reaction $k_j$ involving species $X_{a(j)}$ and $X_{b(j)}$, and $k_{L+1},...,k_{L+N}$ correspond to homodimerisation reactions.
\\\\
We will prove claim (II) for the reaction $X_h+X_v\rightarrow 0$, with reduced propensity $\tilde{f}_j(\vec{\phi}(0))=k_j\phi_i(0)\phi_r(0)$. Note that generally we have $\tilde{f}_j(\Omega,\bold{n'},\vec{\phi}(0))$ but because the reaction involves two abundant species there is no explicit dependence on $\Omega$ and $\bold{n'}$ and hence in what follows we use the less cumbersome notation $\tilde{f}_j(\vec{\phi}(0))$. Using Leibniz's rule,
\begin{align}
\tilde{f}^{(k)}_j(\vec{\phi}(0))=k_j\sum_{l=0}^k \binom{k}{l}\phi_h^{(k-l)}(0)\phi_v^{(l)}(0).
\end{align}
For this reaction, $k_j \propto \frac{1}{x^2}$ and also $\phi_h(0),\phi_v(0) \propto x \rightarrow \infty$. If we can prove that each derivative $\phi_i^{(k)}(0)$ is bounded in the abundant limit, $k>0$, then the above expression for $\tilde{f}^{(k)}_j$ will tend to zero in that limit, owing to the limiting prefactor $k_j$. Firstly, the definition of the first derivative of $\phi_h$ is given by the rate equation:
\begin{align}
\frac{d{\phi}_h(0)}{dt}=&\sum_{j=-1}^{-N}S_{hj}k_j+\sum_{j=1}^N S_{hj}k_j\phi_j(0)+\sum_{j=N+1}^L k_j S_{hj}\phi_{a(j)}(0) \phi_{b(j)}(0)+\sum_{j=L+1}^{L+N}k_jS_{hj}\phi_{j-L}(0)^2.
\end{align}
Wherever we have a concentration which tends to infinity, ($\phi_j(0),~j=M+1,...,N$) it is by definition cancelled out by the corresponding parameter $k_j$. So this expression remains constant, and therefore bounded, in the abundant limit. For the purposes of mathematical induction, assume now that all derivatives up to $\phi_h^{(k)}(0)$ remain bounded in that limit. Then, using Leibniz's rule,
\begin{align}
\phi_h^{(k+1)}(0)=&\sum_{j=1}^N S_{hj}k_j \phi_j^{(k)}(0)+\sum_{j=N+1}^L S_{hj}k_j \sum_{q=0}^k \binom{k}{q} \phi_{a(j)}^{(k-q)}(0)\phi_{b(j)}^{(q)}(0)+\sum_{j=L+1}^{L+N} S_{hj}k_j \sum_{q=0}^k \binom{k}{q} \phi_{j-L}^{(k-q)}(0)\phi_{j-L}^{(q)}(0).
\end{align} 

The only terms here that have a possibility of being unbounded are those involving zeroth derivatives of the $\phi_j$, specifically,
\begin{align}
\sum_{j=N+1}^L S_{hj}k_j \left( \phi_{a(j)}^{(k)}(0)\phi_{b(j)}(0)+\phi_{a(j)}(0)\phi^{(k)}_{b(j)}(0) \right)+2\sum_{j=L+1}^{L+N} S_{hj}k_j  \phi_{j-L}^{(k)}(0)\phi_{j-L}(0).
\end{align}
But these $\phi_j(0)$ will only go to infinity if $j>M$, and in that case the reaction rate $k_j$ will go to zero at least fast enough to counteract the limiting concentrations. So therefore we have shown that each time derivative of the concentrations is bounded in the abundant limit, and therefore each time derivative of the $\tilde{f}_j$ goes to zero in that limit. The proofs for the other reactions are very similar. Therefore we have proved claim (II), and consequently we have proved the convergence of the full marginal CME and the reduced CME for all times, in the abundant limit. 

\section{The accuracy of the reduction for finitely large abundant concentrations and special cases}

The abundant limit as stated previously, is the limit that the rate constants of the reactions removing the abundant species go to zero (in a particular manner) which ensures that the ratios of the abundant to non-abundant concentrations go to infinity. It is in this limit that we have proved that the difference between the reduced and full marginal CME goes to zero. Generally we are interested in the case where the ratio of the abundant to non-abundant concentrations is finitely large, not infinite. In this case the reduced CME is approximate. {Given two identical copies of a chemical system, one placed in a small volume and the other in a much larger volume, and given they have the same finite large ratio of abundant to non-abundant concentrations, we expect the reduced CME to be a better approximation for the system confined in the larger volume. The reason is that the number of molecules of abundant and non-abundant species is larger in the system confined in the larger volume and hence the REs in this case provide a good approximation to the dynamics of the abundant species \cite{grima2010, Ramaswamy2012}, the key principle behind our reduced CME. Another equivalent point of view is that in the larger volume the reactions occur on faster timescales than the reactions in low volumes, due to the larger number of interacting molecules and hence the dynamics of the abundant species in the larger volume are more amenable to being modelled by a continuous approach like the chemical Langevin equation or REs \cite{GillespieCLE}.} 

{Of course, as Gillespie pointed out in Ref. \cite{GillespieCLE}, it is difficult to tell how large should the compartment volume be so that a macroscopic approach becomes a feasible approximation. Hence in this section we explore the accuracy of the reduced CME, for finite ratios of abundant to non-abundant concentrations, by means of stochastic simulations.} In particular we will use the SSA \cite{Gillespie1977} to sample the probability distribution of the full CME \eqref{ME} and the Extrande algorithm \cite{Voliotis2015} to sample the probability distribution of the reduced CME \eqref{apmarcme} for the case where the rate constants of the reactions removing the abundant species scale as $k_j \propto \frac{1}{x^q}$ (where $q$ is the total number of reactant molecules of abundant species involved in reaction $j$) and $x$ is a finite real number. Note that the reduced CME cannot be sampled using the SSA since the propensities are generally time-dependent and hence the use of Extrande. An alternative to the use of the latter algorithm would be to use a method involving the numerical integration of reaction propensities \cite{Anderson2007}. We also show in this section that curiously, for some chemical systems, the exact and approximate marginals are identical even without taking abundance limit.

\subsection{Stochastic simulations}

\subsubsection{Homodimerisation}\label{homodim}
We will investigate an open homodimerisation reaction studied in Ref. \cite{grima2010}:
\begin{align}\label{homodimeq}
\emptyset \xrightleftharpoons[k_3]{k_1} X_1,\quad
X_1+X_1 \xrightarrow{k_2} X_2,\quad X_2 \xrightarrow{k_4} \emptyset.
\end{align}
\begin{figure}[h]
\centering
\begin{tabular}{cc}\includegraphics[scale=0.22]{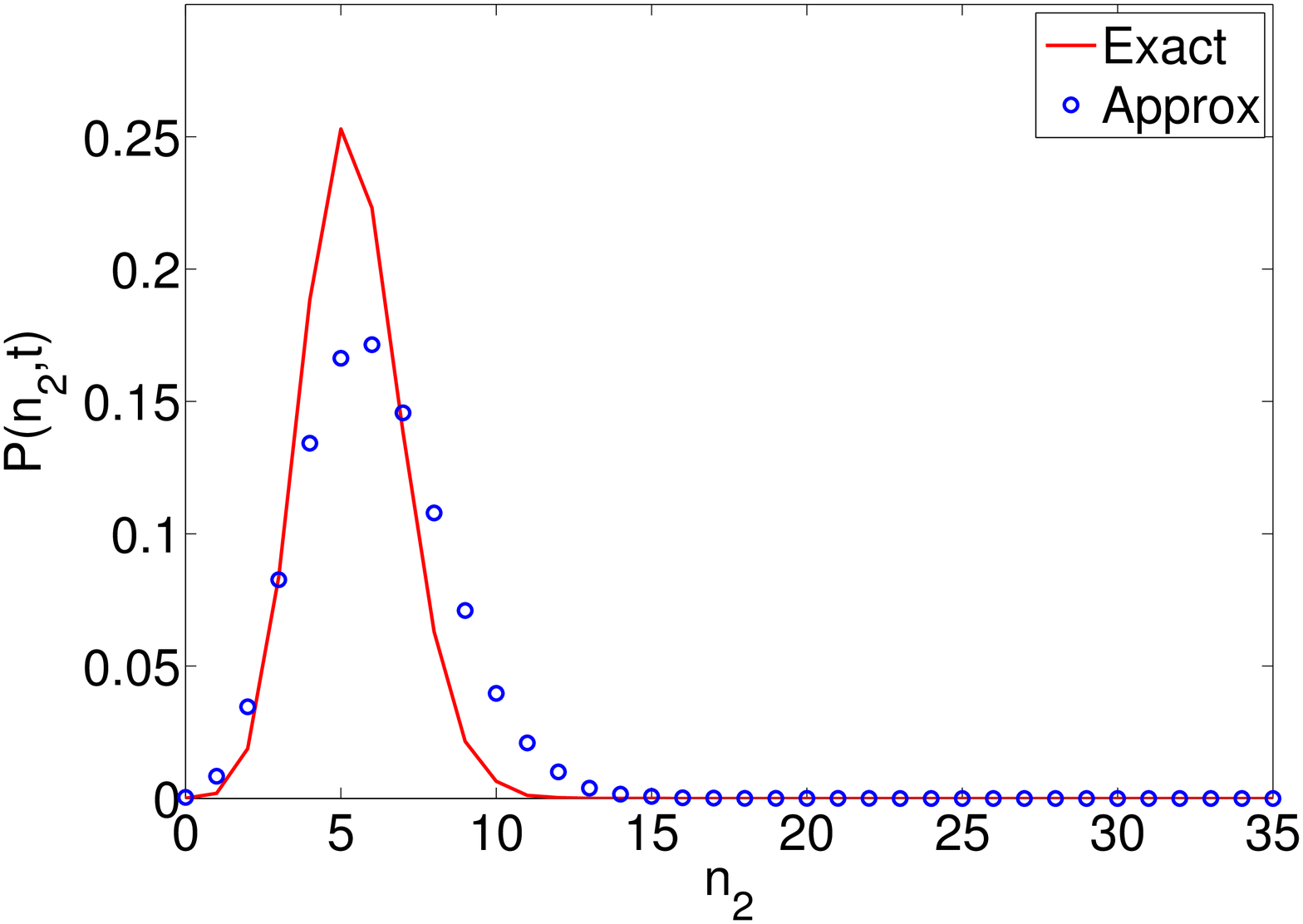} & \includegraphics[scale=0.22]{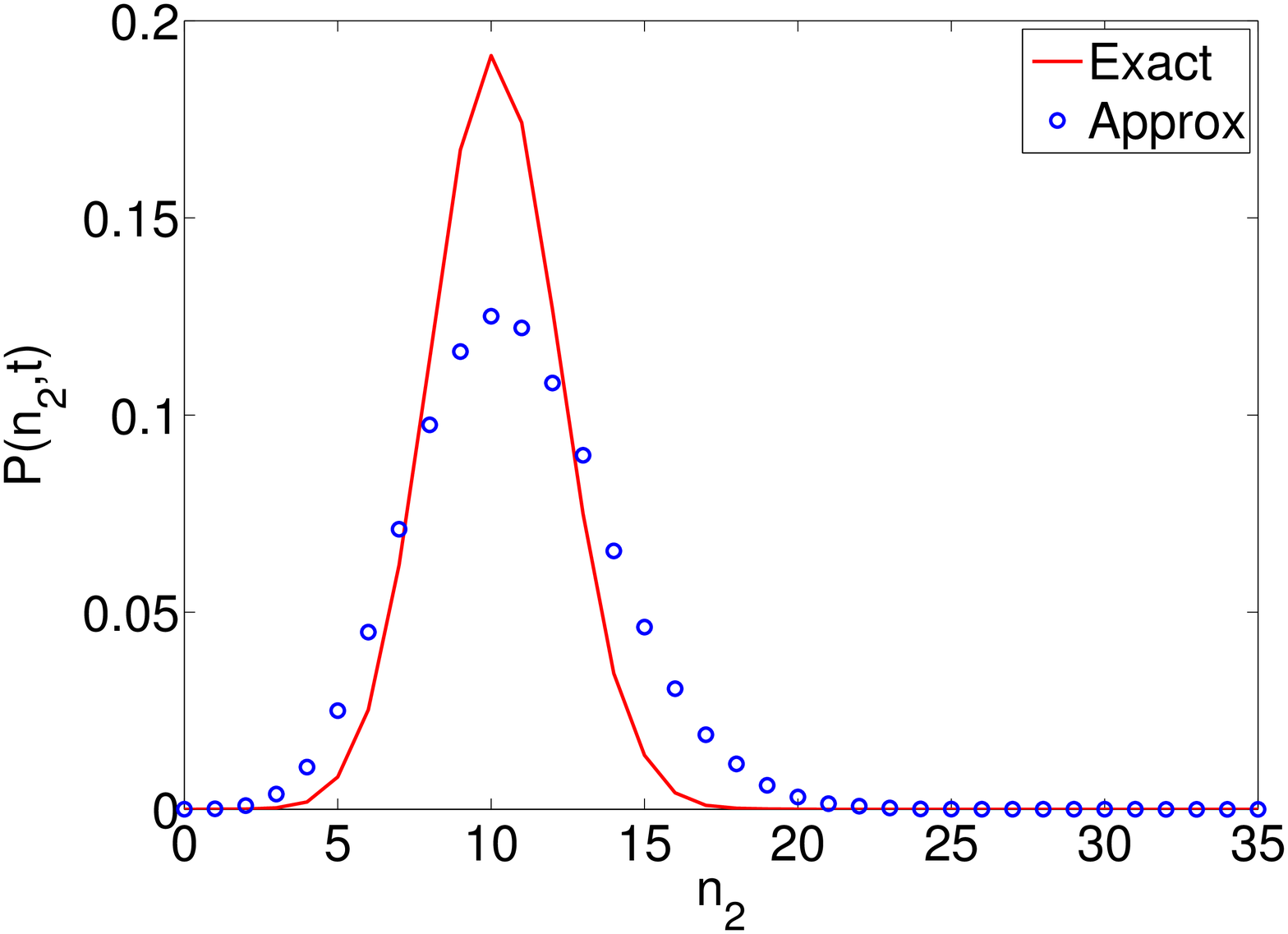}\\
 \small{$\frac{\phi_1(0.01)}{\phi_2(0.01)}=0.4$.} & \small{$\frac{\phi_1(0.01)}{\phi_2(0.01)}=2.7$.}\\
\includegraphics[scale=0.22]{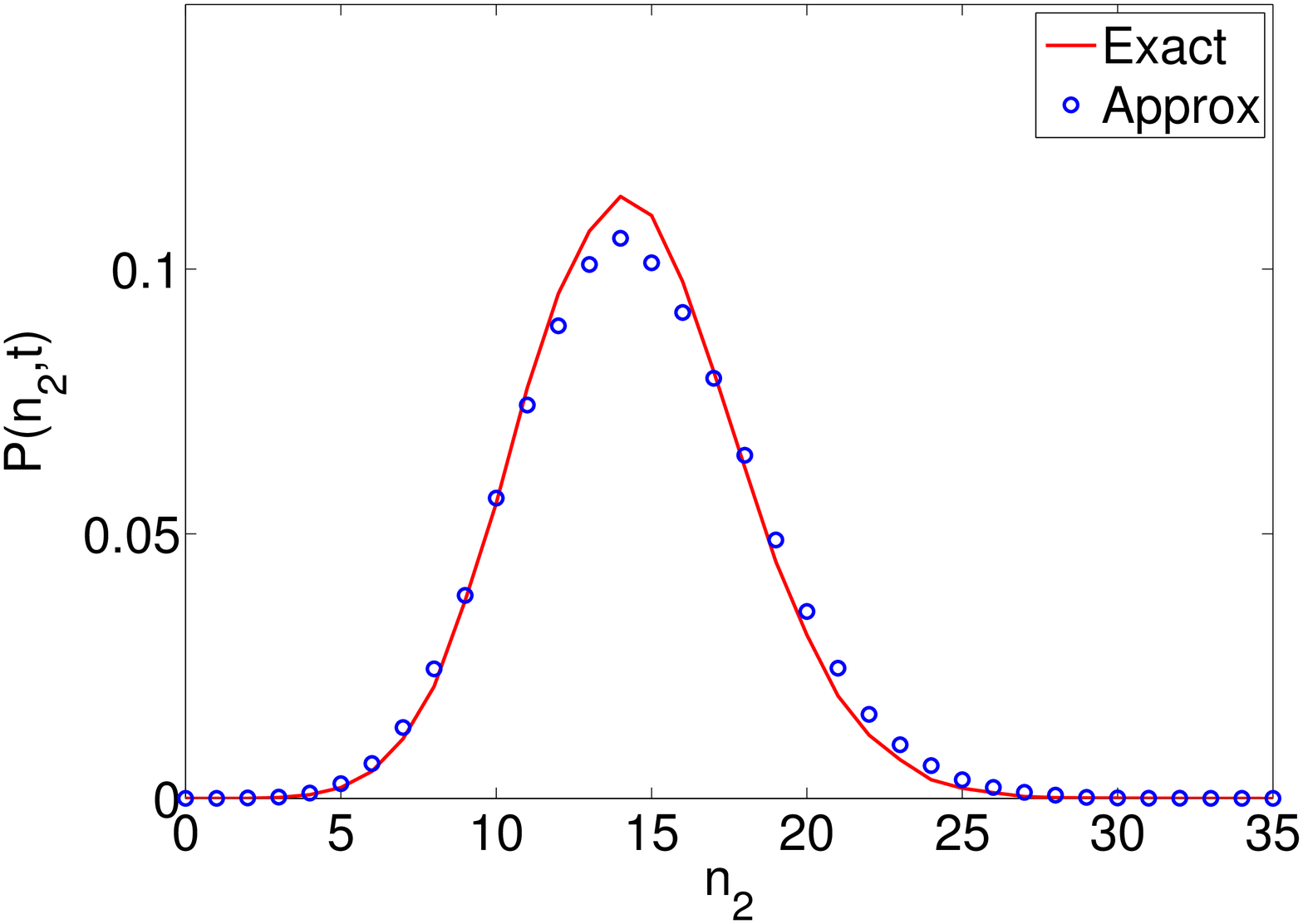}  &\includegraphics[scale=0.22]{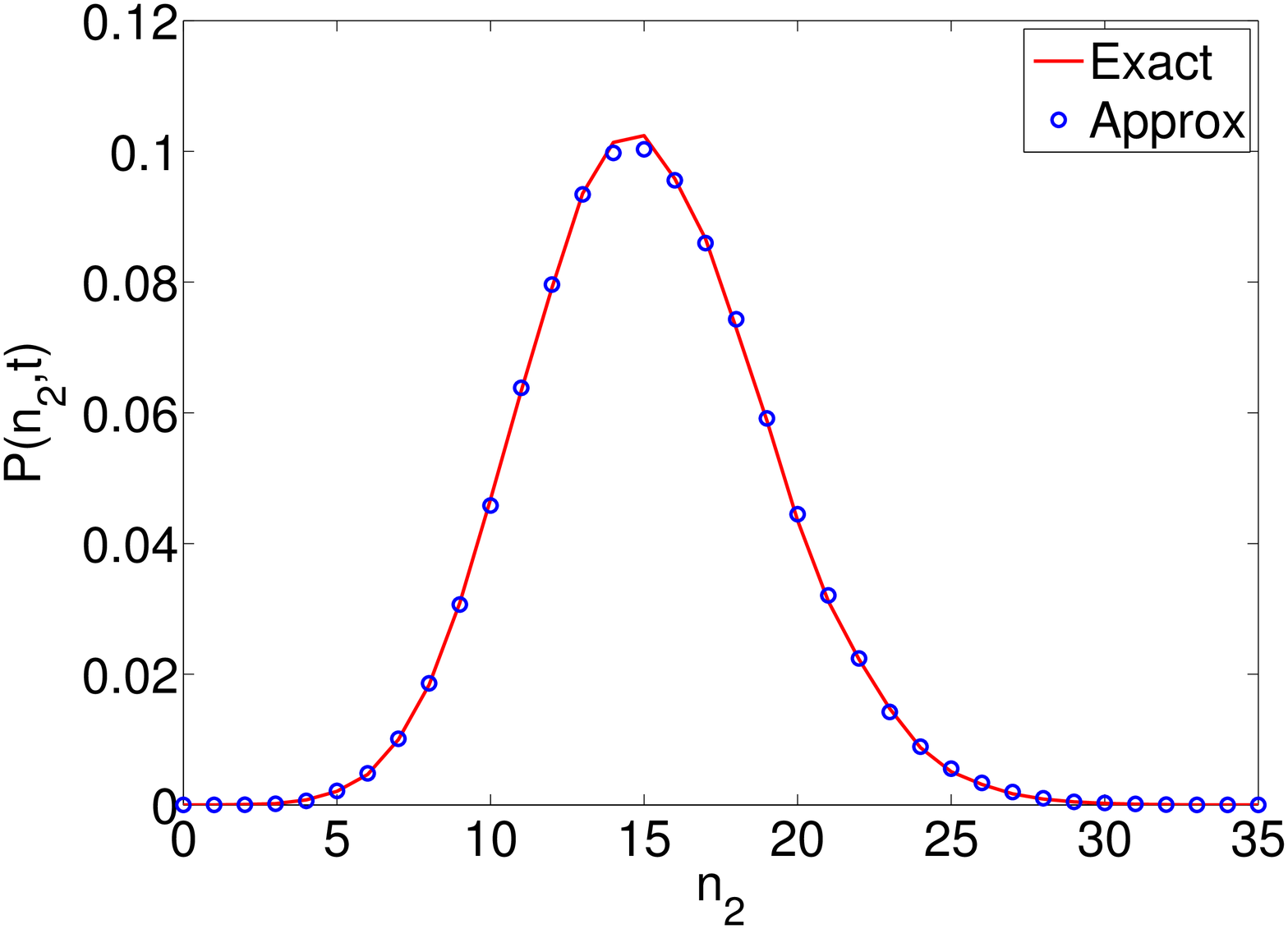}\\
\small{$\frac{\phi_1(0.01)}{\phi_2(0.01)}=26.3$.}&\small{$\frac{\phi_1(0.01)}{\phi_2(0.01)}=262.7$.} \end{tabular}\\
\caption{Exact marginal distribution of species $X_2$ in the open homodimerisation reaction \eqref{homodimeq} (red line) compared with the approximate marginal distribution (blue circles) at time t=0.01, and for different relative abundances $\frac{\phi_1(t)}{\phi_2(t)}$. Parameters are $k_1=10^3$, $k_2=\frac{100}{x^2}$, $k_3=\frac{10}{x}$, $k_4=10$, $\phi_1(0)=4x$, $\phi_2(0)=0$, $\Omega=1$. The parameter $x$ equals $1,10,100,1000$ in (a)-(d), respectively.}\label{fig1}
\end{figure}
Species $X_1$ is produced with rate $k_1$, and is either consumed with rate $k_3$ or else forms a dimer $X_2$ with rate $k_2$. The dimer $X_2$ is then consumed with rate $k_4$. We consider the case where $X_1$ is abundant and $X_2$ is not.
 The RE for the concentration of $X_1$ is:
\begin{align}
\dot{\phi}_1=k_1-2k_2\phi_1^2-k_3\phi_1,
\end{align}
which has the solution:
\begin{align}
\phi_1(t)=-\frac{\alpha \text{tanh} \left( \frac{(\beta-t)\alpha}{2} \right) +k_3}{4k_2},
\end{align}
where
\begin{align}
\alpha = \sqrt{8k_1k_2+k_3^2},\quad
\beta = \frac{2}{\alpha} \text{arctanh} \left( \frac{4k_2 \phi_1(0)+k_3}{-\alpha} \right).
\end{align}
\begin{figure}[h]
\centering
\begin{tabular}{cc} \includegraphics[scale=0.22]{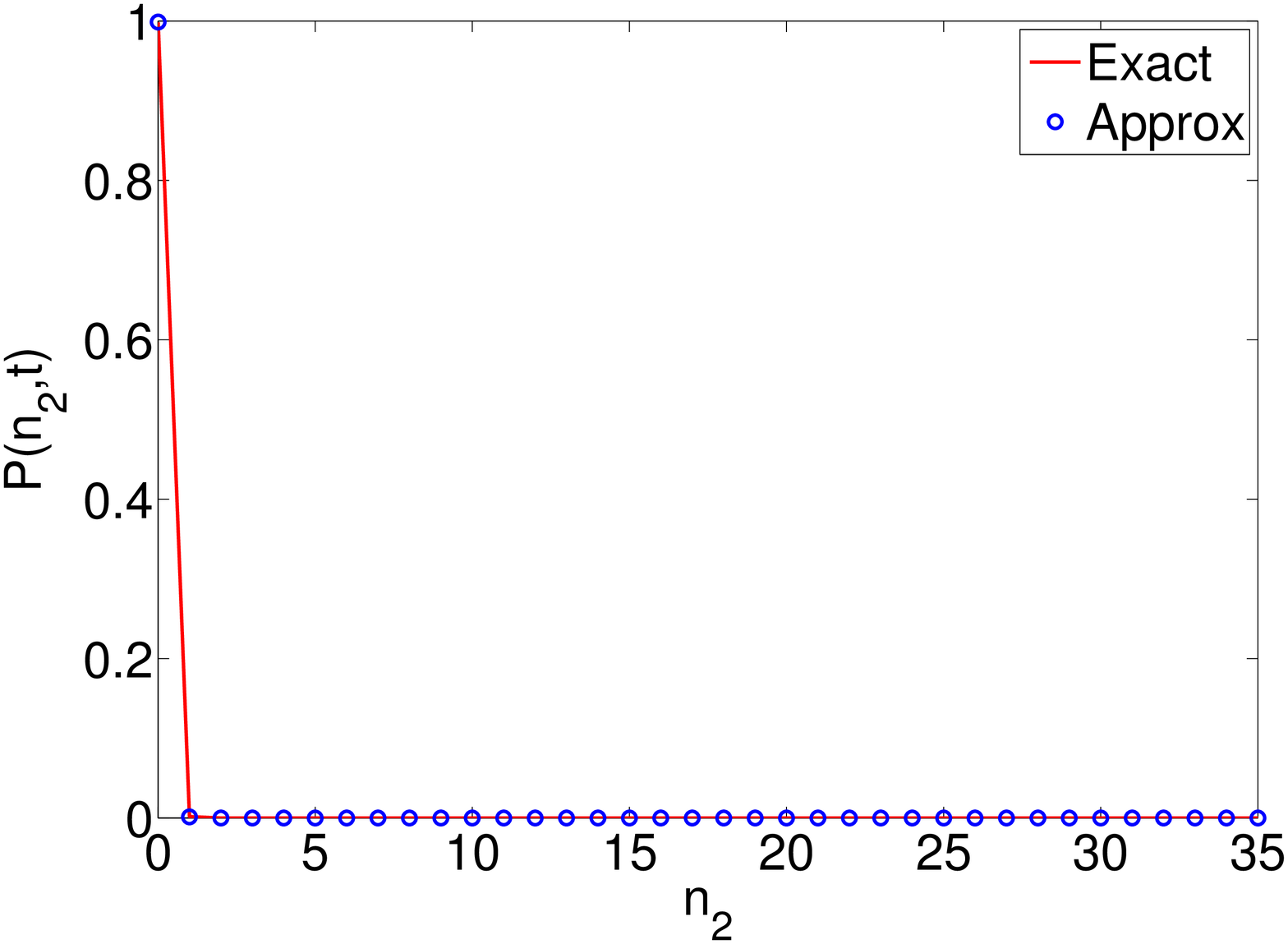} & \includegraphics[scale=0.22]{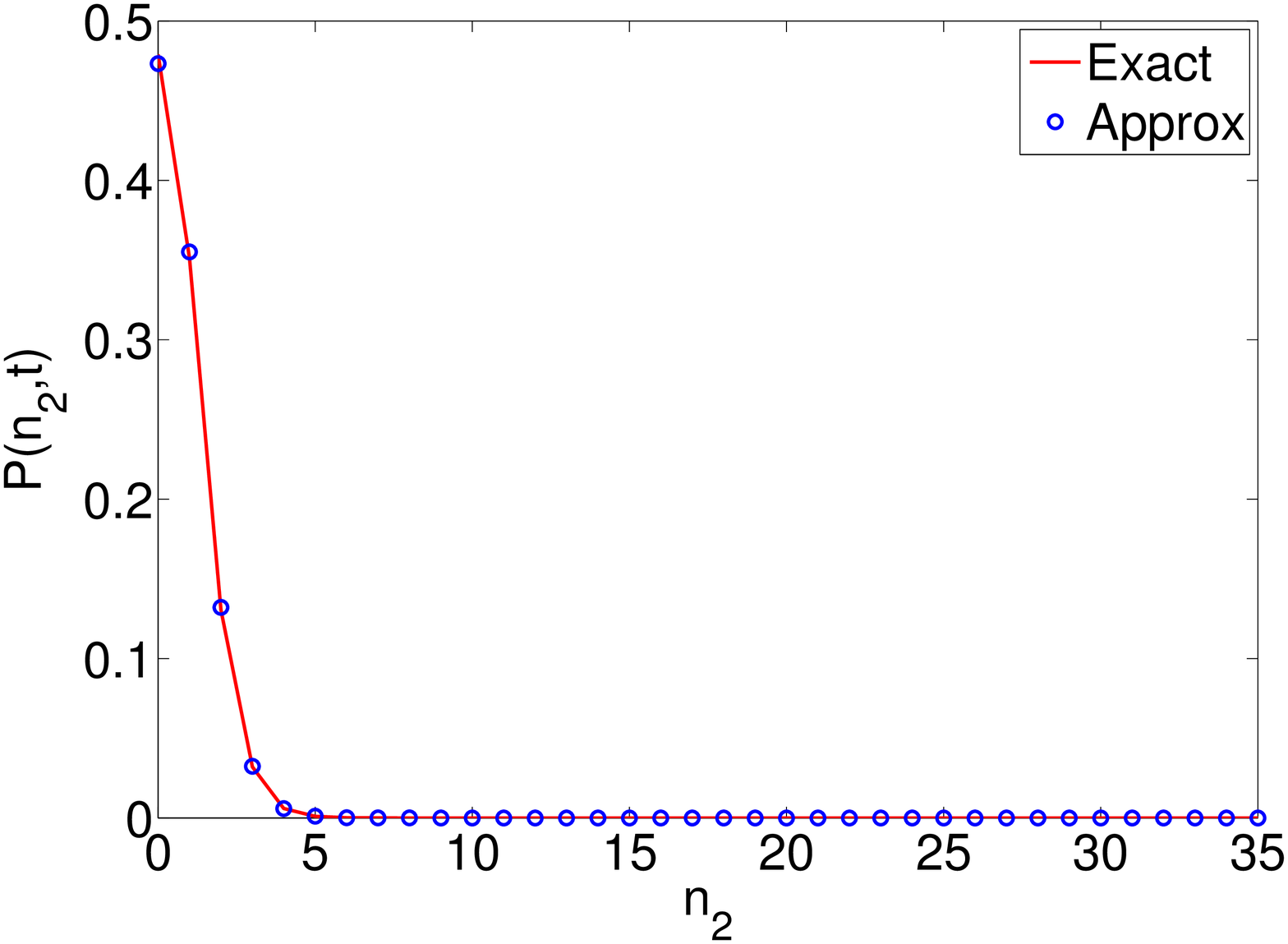}\\
\small{$t=10^{-6}$, $\frac{\phi_1(t)}{\phi_2(t)}=2.5 \times 10^6$.} &\small{$t=5 \times 10^{-4}$, $\frac{\phi_1(t)}{\phi_2(t)}=5.0 \times 10^3$.}\\
\includegraphics[scale=0.22]{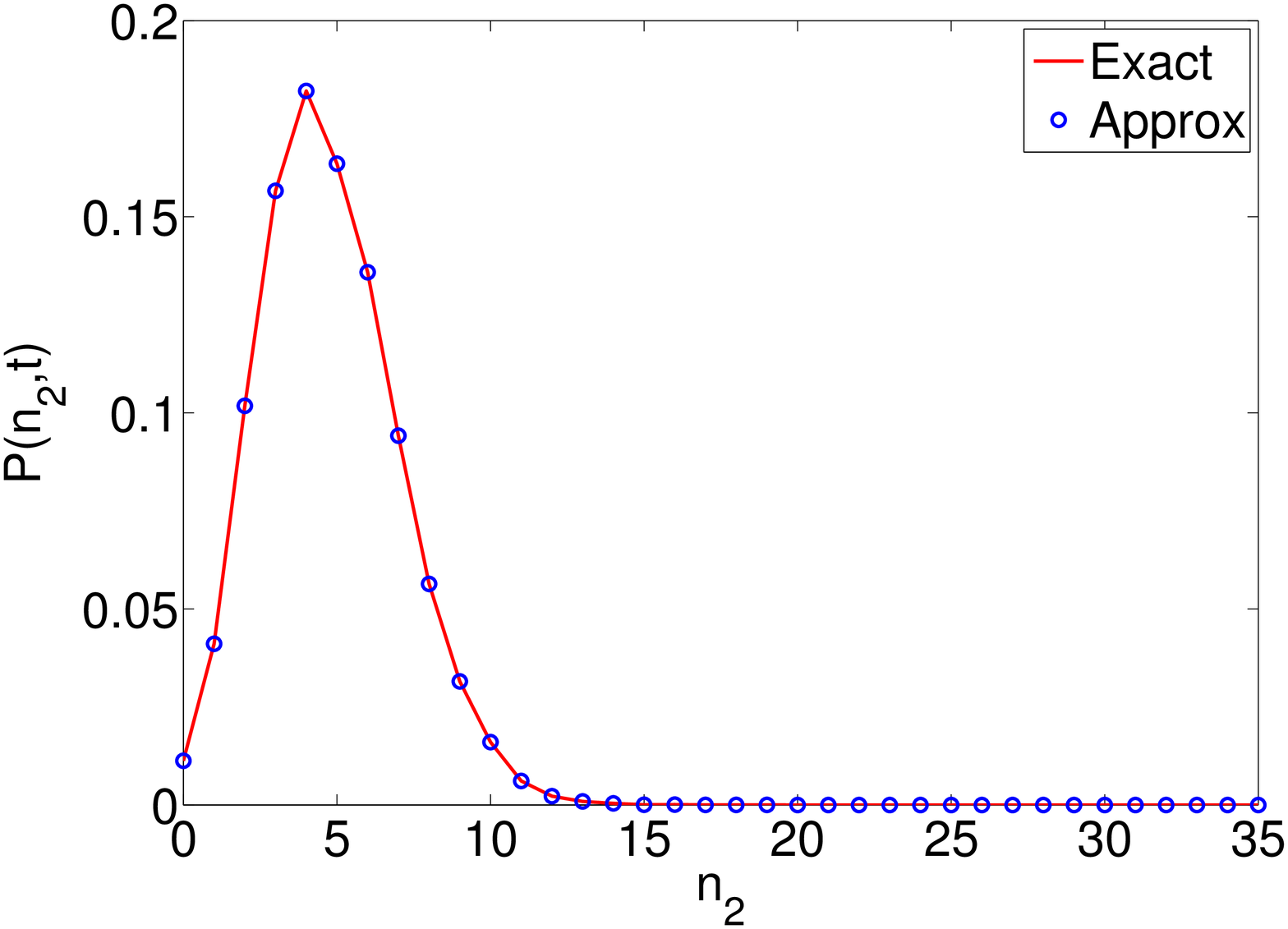} &\includegraphics[scale=0.22]{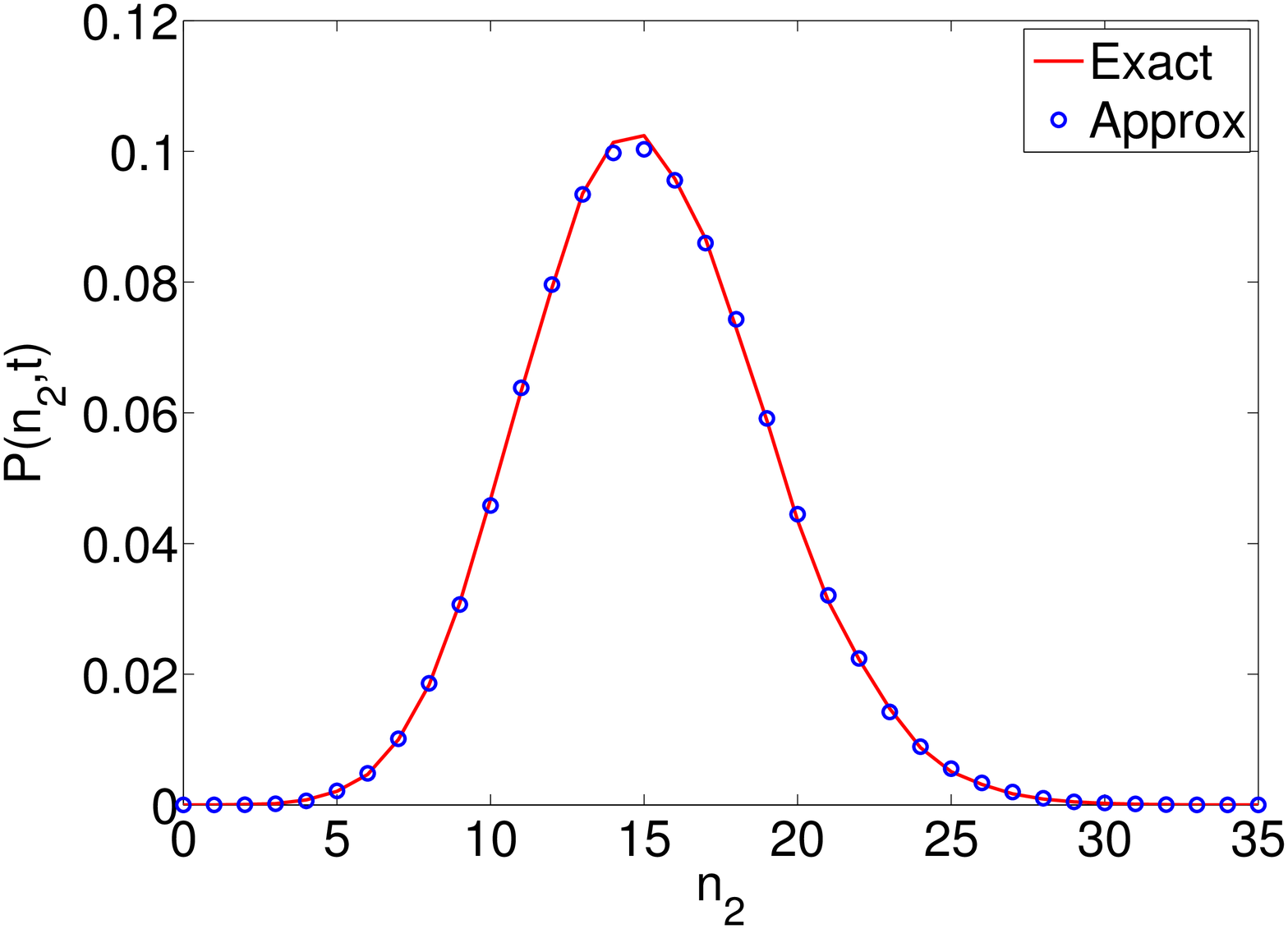}\\
\small{$t=3 \times 10^{-3}$, $\frac{\phi_1(t)}{\phi_2(t)}=8.5 \times 10^2$.}&\small{$t=10^{-2}$, $\frac{\phi_1(t)}{\phi_2(t)}=2.6 \times 10^2$.} \end{tabular}\\
\caption{Time development of the distribution for species $X_2$ in the open homodimerisation reaction \eqref{homodimeq} when abundance is high ($\phi_1(0)=4000$, $\phi_2(0)=0$). There is excellent agreement between exact and approximate marginals at all times. Parameter values are $k_1=10^3$, $k_2=10^{-4},~k_3=10^{-2}$, $k_4=10$, $\Omega=1$.}\label{fig2}
\end{figure}
Therefore the approximate reduced CME Eq. \eqref{apmarcme} for non-abundant species $X_2$ is given by:
\begin{align}
\dot{\tilde{P}}(n_2,t)&=\Omega k_2 \phi_1^2(t) \left( \tilde{P}(n_2-1,t)-\tilde{P}(n_2,t) \right)        +k_4 \left( (n_2+1)\tilde{P}(n_2+1,t)-n_2\tilde{P}(n_2,t) \right).
\end{align}
In Figs. \ref{fig1} and \ref{fig2} we compare the exact marginal distribution of $X_2$ from the full CME and the approximate marginal distributions of the reduced CME. These are obtained by means of an ensemble average of SSA and Extrande trajectories, respectively. In Fig. \ref{fig1} we plot the distributions for four different relative abundances of $X_1$ and $X_2$ at the same time point $t=0.01$. The abundance is adjusted by choosing the rate constants to scale as $k_2 \propto \frac{1}{x^2}$ and $k_3 \propto \frac{1}{x}$ (in accordance with the limits delineated in Section 2.2) and varying $x$ over a certain finite range (see caption of Fig. 1). The distance between the two distributions clearly becomes smaller as the ratio of the abundant to non-abundant species  concentrations increases, in line with the proof of the previous section; the two distributions are practically indistinguishable when this ratio is of the order of 100. Nevertheless we find that some salient features of the two distributions are fairly similar (namely the position of mode and the width of distribution) over a large range of the ratio of the abundant to non-abundant species concentrations (0.4 to to 263). In Fig. \ref{fig2} we show that the high accuracy of the reduced CME also extends to predicting the whole time-evolution of the distribution. Both of these figures indicate that the results of simulations using the reduced CME bear a significant closeness to those obtained using the full CME under a wide range of abundances and hence point towards the utility of the reduced CME as a low dimensional approximation of the full CME.

\subsubsection{Genetic Feedback Loop}\label{genfeed}
We next investigate a negative genetic feedback loop studied in Ref. \cite{grima2012steady}.
\begin{align}\label{genfeedeq}
D_{\text{on}}\xrightarrow{v_0} D_{\text{on}}+P,\quad
P \xrightarrow{d_0} \emptyset,\quad
D_{\text{off}} \xrightleftharpoons[d_1]{v_1} D_{\text{on}}+P.
\end{align}
The ``on" promoter $D_\text{on}$ produces proteins $P$ with rate $v_0$. These proteins bind to the promoter with rate $d_1$ to generate the inactive ``off" promoter $D_\text{off}$, which can then unbind back into active promoter and protein with rate $v_1$. Furthermore, the protein $P$ is consumed by a unimolecular reaction with rate $d_0$. We consider the case where $P$ is abundant.

The REs for this system are given by:
\begin{align}
\dot{\phi}_1&=-v_1\phi_1+d_1\Big(\frac{1}{\Omega}-\phi_1\Big)\phi_2,\nonumber\\
\dot{\phi}_2&=v_1\phi_1-d_1\Big(\frac{1}{\Omega}-\phi_1\Big)\phi_2+v_0\Big(\frac{1}{\Omega}-\phi_1\Big)-d_0\phi_2,
\end{align}
where $\phi_1$ is the concentration of $D_{\text{off}}$, $\phi_2$ is the concentration of $P$, $\Omega$ is the volume and $\frac{1}{\Omega}$ is the total (fixed) gene concentration (equivalent to one gene).
\begin{figure}[h]
\includegraphics[scale=0.41]{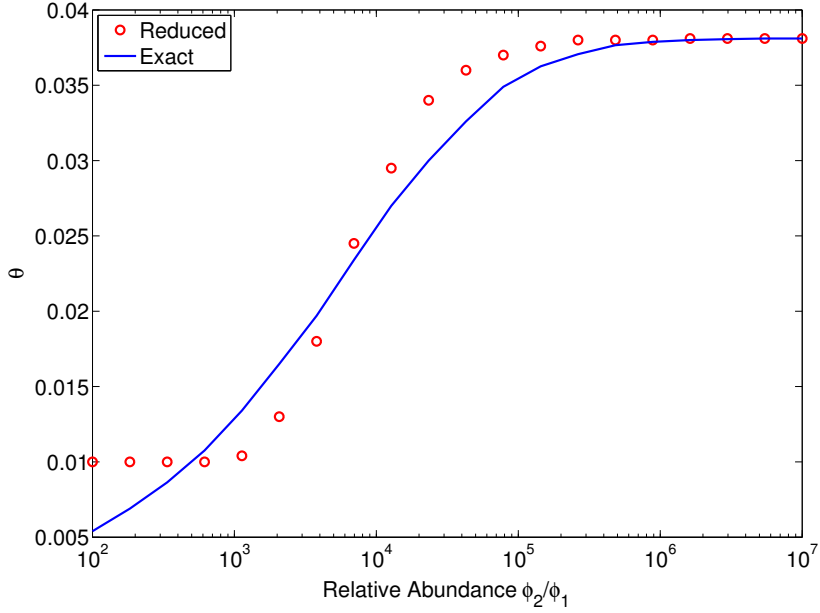}
\caption{The parameter $\theta=\langle n_1 \rangle$ for the genetic feedback loop \eqref{genfeedeq}. The exact value of the parameter is obtained from an ensemble average of SSA trajectories of the full CME (blue line), while the approximate value is that obtained from our reduced CME via numerical solution of Eq. \eqref{thetaeq} (red circles). The time is fixed to $t = 50$. Parameter values are $\Omega=1,~v_0=1,~v_1=0.1,~d_0=\frac{1}{x},~d_1=\frac{0.01}{x},~1<x<10^4$.}
\end{figure}
Unlike the previous example, these equations do not admit an analytical solution, so we must solve them numerically, and then use the numerical solution in the reduced CME:
\begin{align}
\dot{\tilde{P}}(n_1,t)&=v_1\left((n_1+1)\tilde{P}(n_1+1,t)-n\tilde{P}(n_1,t) \right)+d_1\phi_2(t) \left( (2-n_1)\tilde{P}(n_1-1,t)-(1-n_1)\tilde{P}(n_1,t) \right),
\end{align}
where $n_1$ is the number of molecules of $D_{\text{off}}$.
Since at any one time, the gene is either on or off, the distribution of $D_\text{off}$ is Bernoulli. In particular, since the parameter of the Bernoulli distribution, $\theta(t)$, is equal to $\tilde{P}(1,t)$, we can say,
\begin{align}
D_\text{off} \sim \text{Bernoulli}(\theta(t)),
\end{align}
where $\dot{\theta}(t)$ is obtained by setting $n_1=1$ in the above reduced CME,
\begin{align}\label{thetaeq}
\dot{\theta}=-v_1\theta(t)+d_1\phi_2(t) \left( 1-\theta(t) \right).
\end{align}
Our reduction method therefore provides us with an ``exact" solution at all times in this case, since we don't need to perform any stochastic simulations sampling the reduced CME, but rather just numerically solve the ordinary differential equation above.

Figure 3 shows how the approximate expression for $\theta$ given in Eq. \eqref{thetaeq} compares with the true $\theta$ (obtained by computing $\langle n_1 \rangle$ from an ensemble average of SSA trajectories of the full CME) for different relative abundances at $t=50$ seconds. The relative abundance is controlled by choosing the rate constants to scale as $d_0, d_1 \propto \frac{1}{x}$ (in accordance with the limits delineated in Section 2.2) and varying $x$ over a certain finite range (see caption of Fig. 3). For the parameter set chosen, we find that the approximation for $\theta$ using the reduced CME is in good qualitative agreement with that calculated from the full CME when the ratio of abundant to non-abundant concentrations varies over the range $10^3-10^7$. In particular both the full and reduced CME predict that the probability of the gene being in the off state increases monotonically, in a step-like manner, as the protein concentration increases at constant gene concentration (consistently with a negative feedback loop). For low relative abundances (ratios less than a few hundreds), the approximate $\theta$ is almost double the true value implying that the reduced CME in this case over-estimates the strength of the negative feedback.  

{\subsubsection{Metabolic network}
We consider an arbitrarily large, sequential enzyme reaction network which has been previously associated with metabolism \cite{tschudy1973steady,thomas2010stochastic}. The network consists of $N+1$ enzymes, each converting a substrate into a product which then serves as the substrate for the next enzyme in the cascade. The first substrate is produced by a zeroth-order reaction, and the final substrate is converted into a product which we ignore. We seek the approximate distribution of the enzymes, which we expect to be exact in the limit where substrates are abundant. 
The full chemical system is described by the scheme:
\begin{align}\label{MetNet}
\emptyset &\xrightarrow{k_\text{in}} S_0,\nonumber\\
E_0+S_0 &\xrightleftharpoons[k_{-1}^0]{k_1^0}C_0\xrightarrow{k_2^0}E_0+S_1,\nonumber\\
E_1+S_1 &\xrightleftharpoons[k_{-1}^1]{k_1^1}C_1\xrightarrow{k_2^1}E_1+S_2,\\
&~~~\vdots\nonumber\\
E_N+S_N &\xrightleftharpoons[k_{-1}^N]{k_1^N}C_N\xrightarrow{k_2^N}E_N.\nonumber
\end{align}
\begin{figure}[h]
\centering
\begin{tabular}{cc} \includegraphics[scale=0.21]{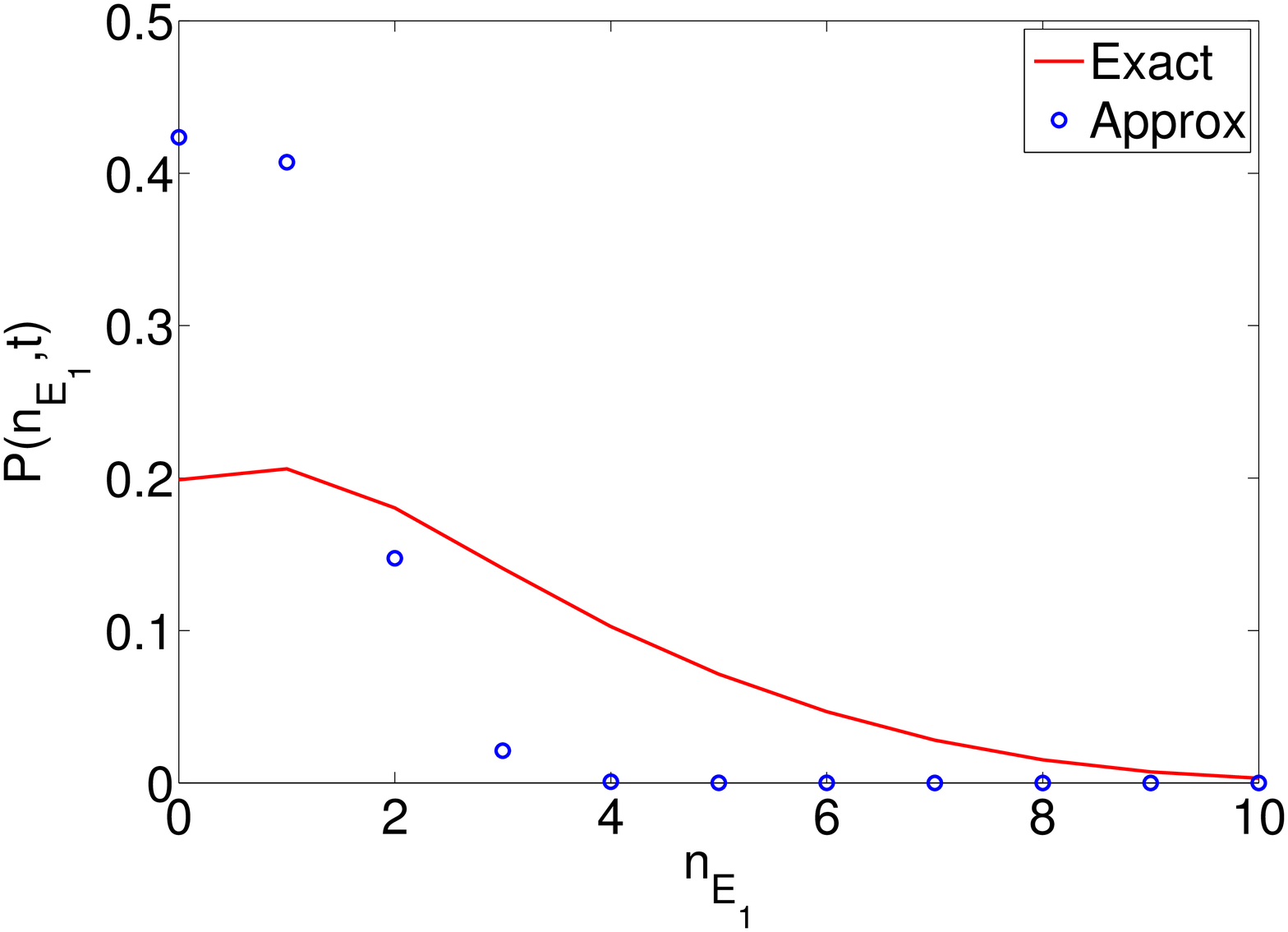} & \includegraphics[scale=0.21]{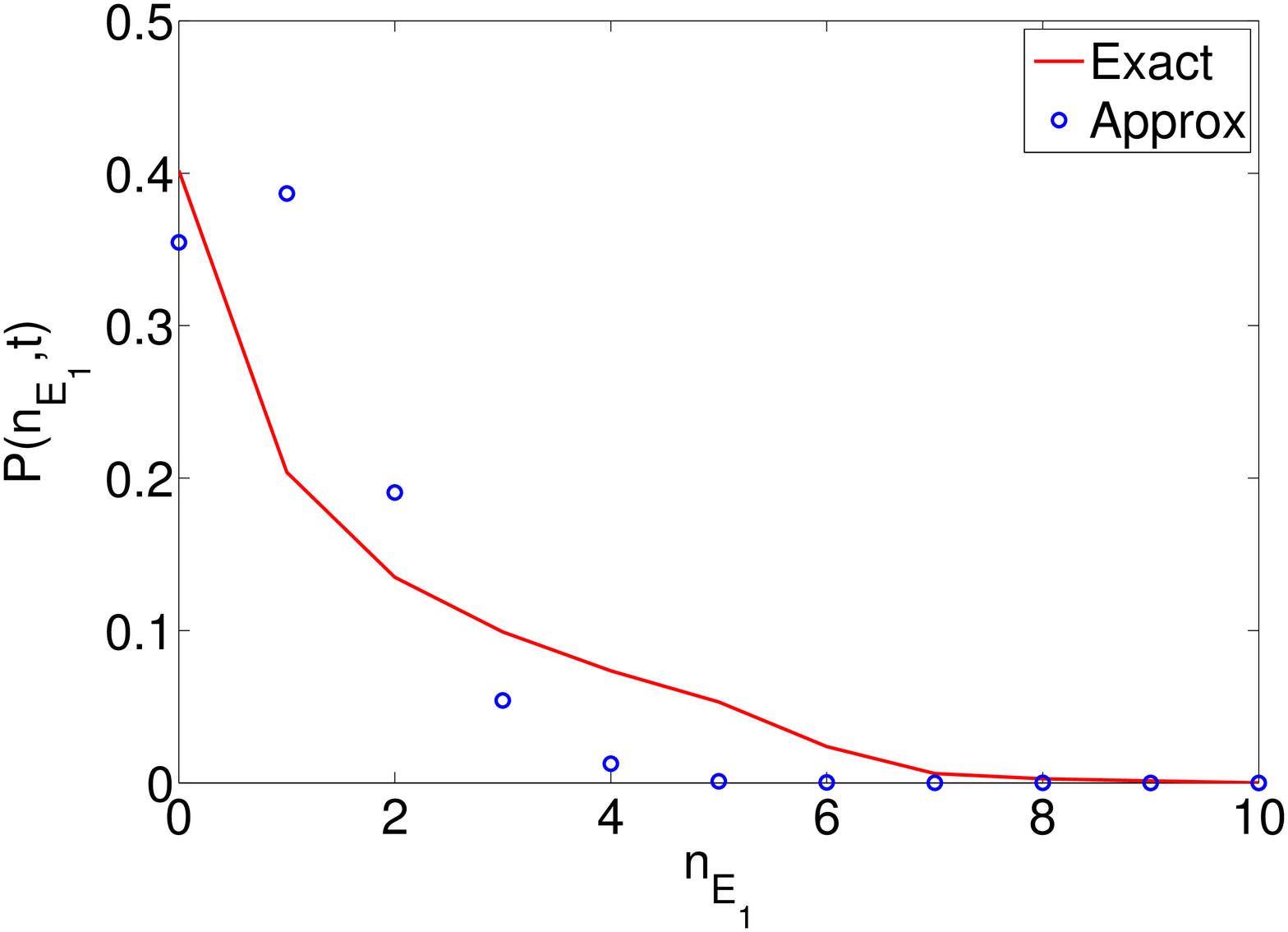}\\
\small{$k_1^i=135$, $\Omega \phi_{S_1}(t)/D_1=0.01$.} &\small{$k_1^i=13.5$, $\Omega \phi_{S_1}(t)/D_1=0.1$.}\\
\includegraphics[scale=0.21]{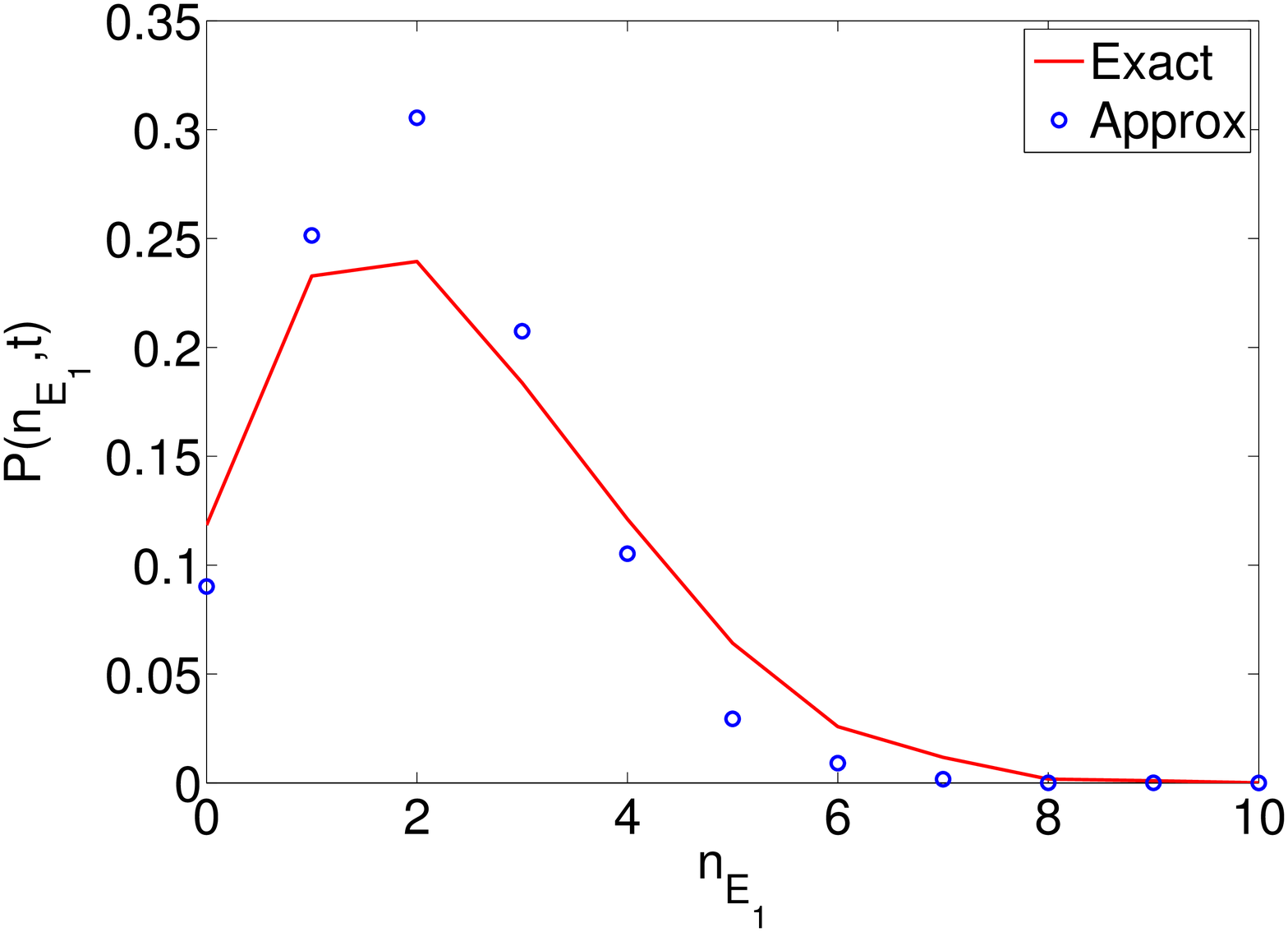} &\includegraphics[scale=0.21]{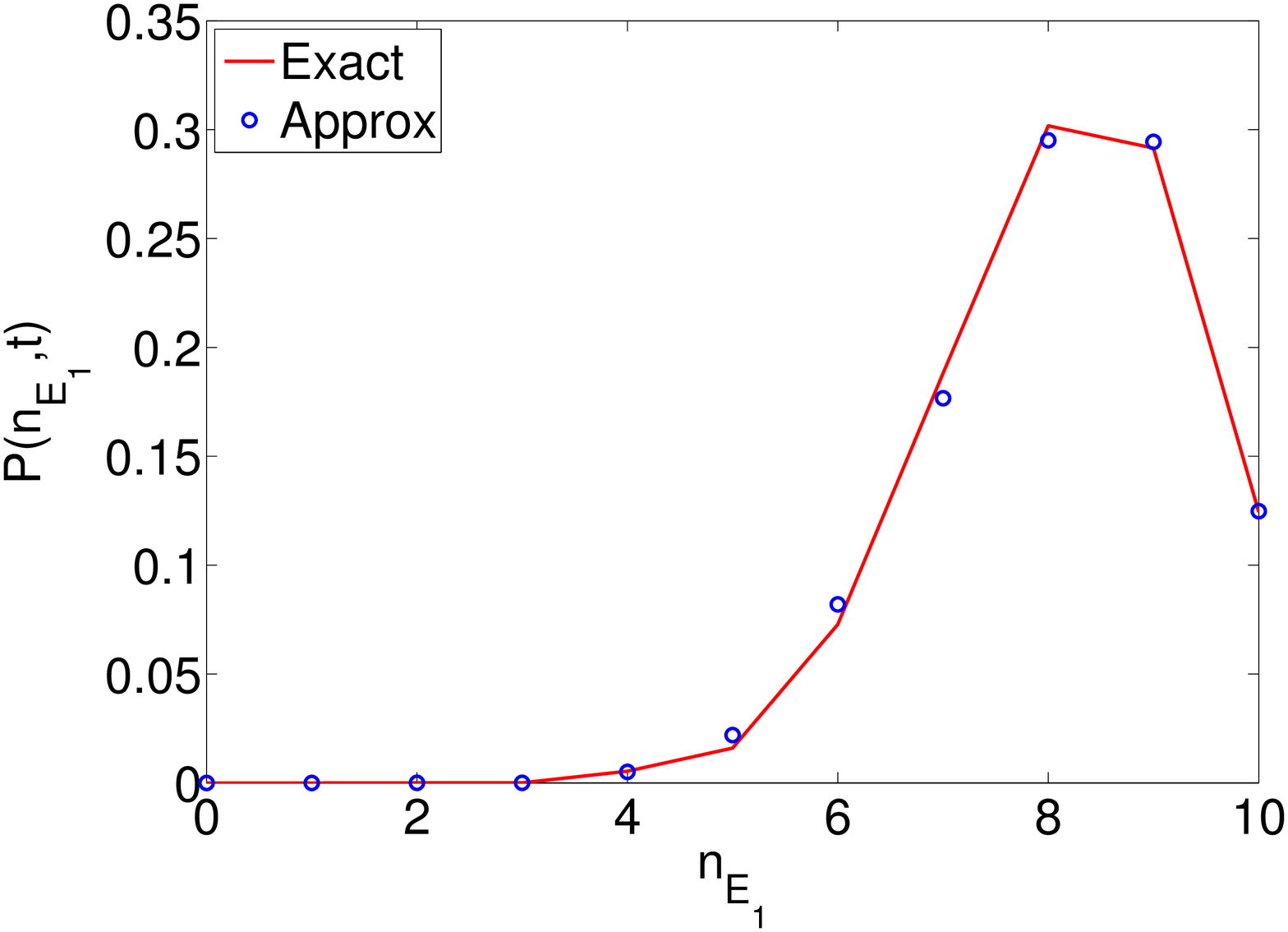}\\
\small{$k_1^i=1.21$, $\Omega \phi_{S_1}(t)/D_1=1$.}&\small{$k_1^i=0.02$, $\Omega \phi_{S_1}(t)/D_1=10$.} \end{tabular}\\
\caption{{Probability distribution of the number of molecules of species $E_1$ in the metabolic network model \eqref{MetNet}, for different substrate to enzyme abundance ratios: 0.1, 1, 10, and 100  respectively. This system consists of 11 distinct enzyme and 11 substrate species, though here we only look at the distribution of $E_1$ relative to the abundance of $S_1$, i.e, $\Omega \phi_{S_1}(t)/D_1$. There is clear convergence between the approximate and exact distributions as the substrate becomes more abundant than the enzyme. Parameters are $\Omega=1$, $t=100$, $N=10$, $D_i=10$, $k_\text{in}=18$, $k_{-1}^i=1$, $k_2^i=2$, $i=0,...,N$.}}\label{revplot1}
\end{figure}
The conservation law $n_{E_i}+n_{C_i}=D_i$ is implied for each enzyme, where $D_i$ is a given positive integer which represents the total number of both free and bound enzymes of type $i$, which is constant in time. According to our method, encapsulated by Eq. (\ref{redchemsysN}), the reduced chemical system takes the simpler form:
\begin{align}
\label{redchemsysmetN}
E_i &\xrightleftharpoons[k_{-1}^i+k_2^i]{k_1^i\phi_{S_{i}(t)}}C_i,~~i=0,...,N.
\end{align}
It is hence clear that the molecule numbers of each enzyme species are binomially distributed in steady-state conditions:
\begin{equation}
\label{ssMetNet}
\tilde{P}(n_{E_i})=\binom{D_i}{n_{E_i}}\left( \frac{k_{-1}^i+k_2^i}{k_1^i\phi_{S_i}+k_{-1}^i+k_2^i}\right)^{n_{E_i}}\left( \frac{k_1^i\phi_{S_i}}{k_1^i\phi_{S_i}+k_{-1}^i+k_2^i}\right)^{D_i-n_{E_i}},
\end{equation}
where $\phi_{S_i}$ is the steady-state solution of the REs of the full system, Eqs. (\ref{MetNet}), given by:
\begin{equation}
\label{ssMetNetD}
\phi_{S_i}=\frac{k_\text{in}\left(k_2^i+k_{-1}^i\right)}{k_1^i \left( \frac{k_2^iD_i}{\Omega}-k_{\text{in}}\right)}.
\end{equation}}

{For a time-dependent description, the reduced CME corresponding to the reduced chemical system (\ref{redchemsysmetN}) cannot be exactly solved and stochastic simulations are required. In Fig. \ref{revplot1} we plot both the approximate and exact distributions (using Extrande for the reduced system and the SSA for the full system) of the enzyme $E_1$ at a fixed time for different abundances of substrate $S_1$. It is clear that the approximation improves as the substrate becomes more abundant than enzyme, and is essentially exact in the bottom right panel where the relative abundance is $\Omega \frac{\phi_{S_1}}{D_1} = 10$. It is also remarkable that the approximation is good even when there is essentially no clear separation in abundance, i.e,  $\frac{\Omega \phi_{S_1}}{D_1} = 1$. Indeed even though the approximation suffers quantitatively when the relative abundance is not high, yet it captures the main distinctive qualitative feature, namely that the distribution changes from positive to negative skewness as a function of the relative abundance (the switch happens at a relative abundance between 1 and 10).
\begin{figure}[h]
\centering
\includegraphics[scale=0.38]{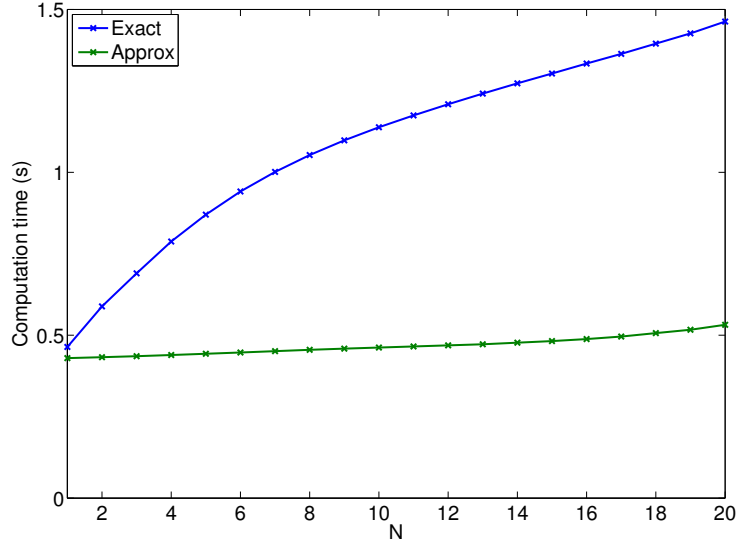}
\caption{{Computational time taken to compute an individual trajectory of length $100$ time units for species $E_1$ in the metabolic network model  with a total of $2N+2$ species, using the Exrande algorithm for the reduced (approximate) chemical system (\ref{redchemsysmetN}) and the SSA for the full (exact) chemical system (\ref{MetNet}). Parameters are $\Omega=1$, $D_i=10$, $k_\text{in}=18$, $k_1^i=0.02$, $k_{-1}^i=1$, $k_2^i=2$ $\forall i$. Simulations were performed using MATLAB on a computer with a 3GHz Intel Core 2 Duo processor and 4GB RAM. Note that as the total number of species increases, stochastic simulations of the reduced system (approximate) becomes significantly more computationally efficient than the SSA for the full system (exact).}}\label{revplot2}
\end{figure}}

{For this system we have the added benefit that the distribution of the number of molecules of each enzyme $E_i$ is independent in the approximate description. This means that if we are interested in the distribution of the number of molecules of a given enzyme, say, $E_1$, then we only need to simulate the three reactions involving that particular enzyme, rather than the $3N+4$ reactions of the full system. There is therefore a marked reduction in computational time for our reduced SSA, particularly for large $N$, as shown in Fig. \ref{revplot2}, where the approximate SSA is roughly 3 times faster than the exact SSA when $N=20$. We note, however, that the computational time of the approximate method does increase slightly with $N$, owing to the need to solve a coupled system of $2N+2$ rate equations.}

{
\subsubsection{Genetic oscillator with transcriptional feedback}
\begin{figure}[h]
\centering
\begin{tabular}{cc} \includegraphics[scale=0.22]{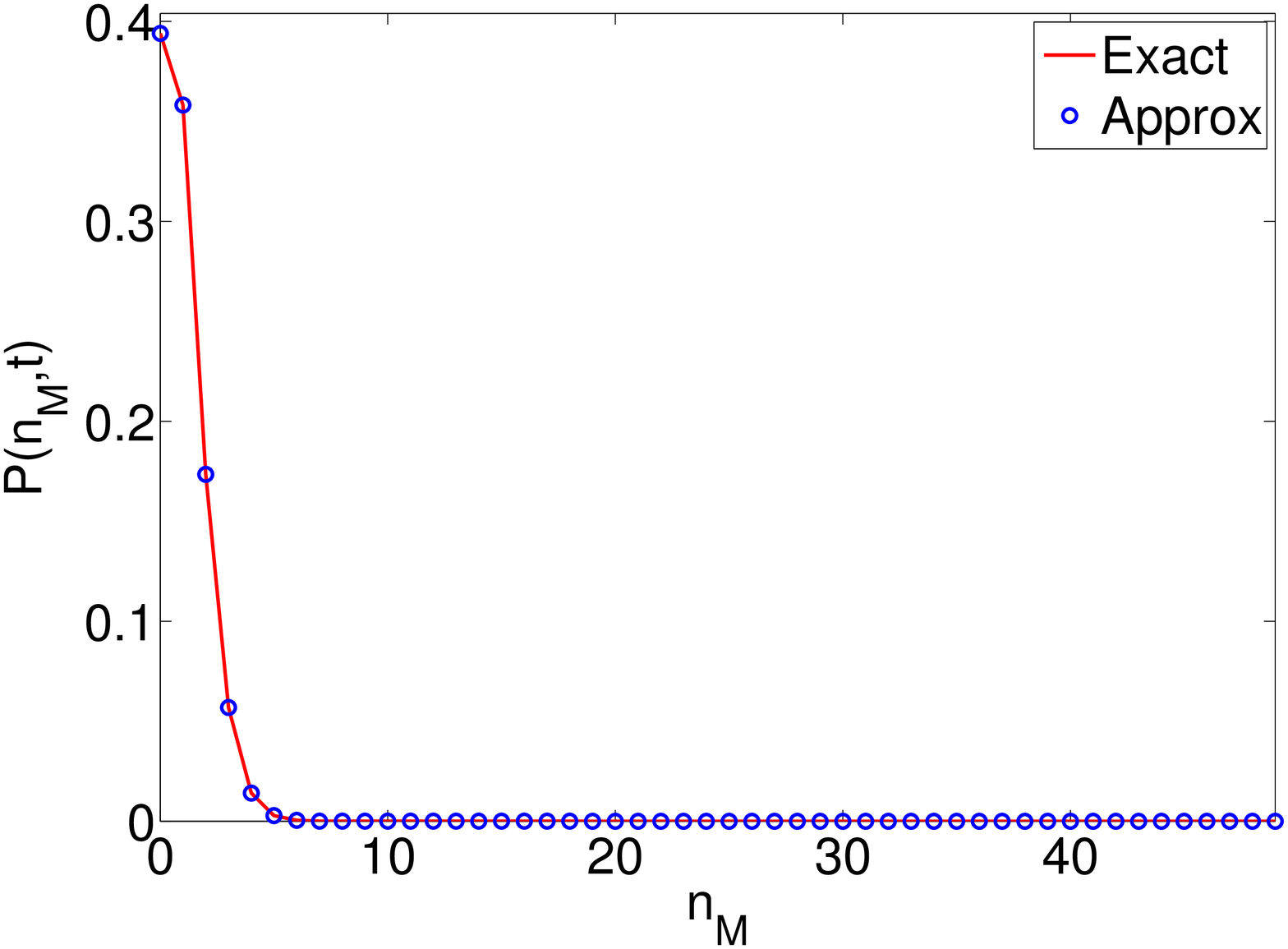} & \includegraphics[scale=0.22]{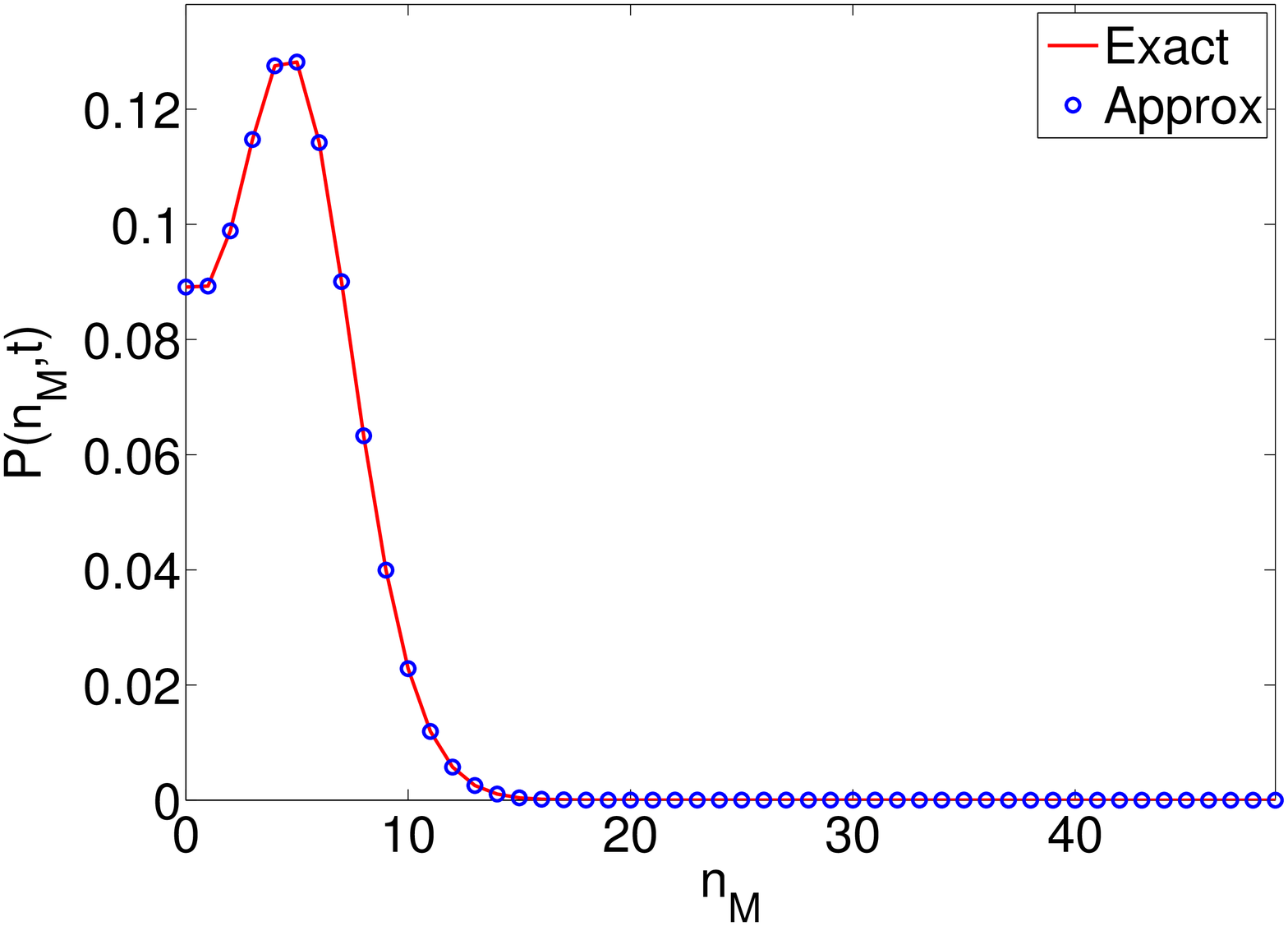}\\
\small{$t=0.01$.} &\small{$t=0.06$.}\\
\includegraphics[scale=0.22]{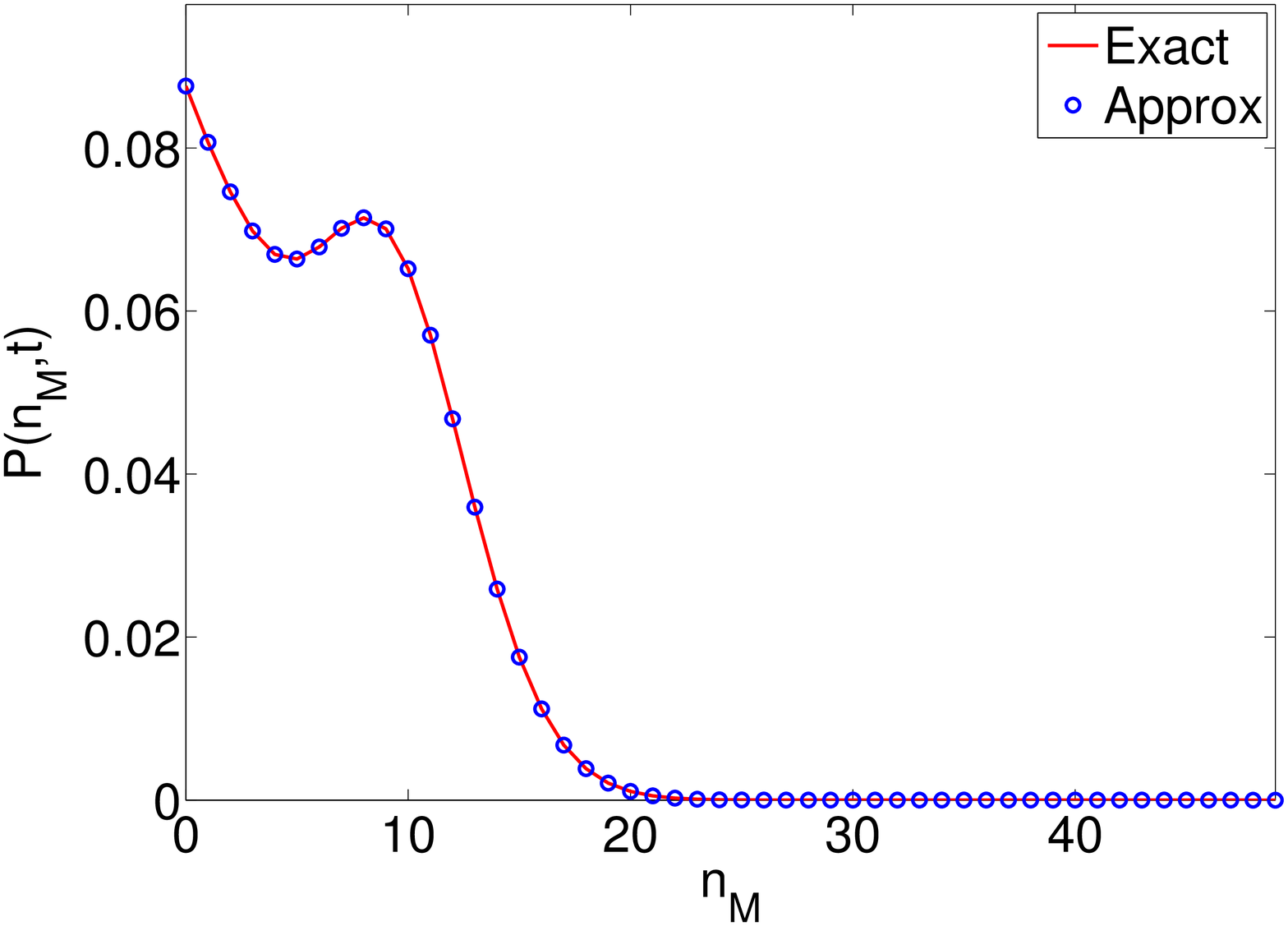} &\includegraphics[scale=0.22]{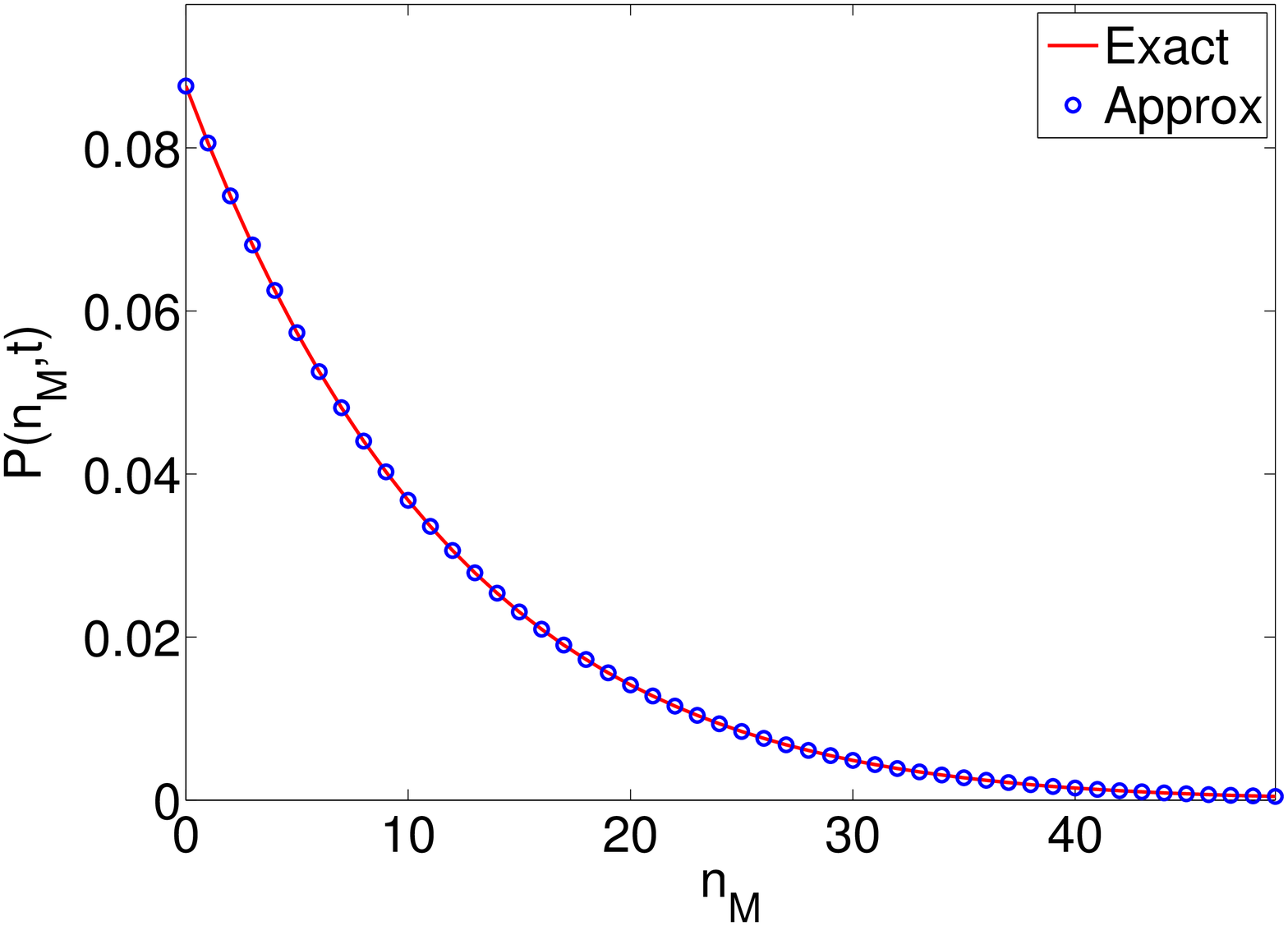}\\
\small{$t=0.11$.}&\small{$t=1$.} \end{tabular}\\
\caption{{Distribution of the number of mRNA, $M$, in the genetic oscillatior model \eqref{GenOsc}, for parameter values which give the steady-state abundance $\frac{\phi_{P_N}}{\phi_M}=100$ and at different times: 0.01, 0.06, 0.11, and 1 seconds respectively. The exact and approximate distributions agree for all times. Remarkably, the approximate system also shows the bimodal behaviour characteristic of the full system, as shown by the bimodal distribution at $t=0.11$. Parameters are $N=10$, $\Omega=1$, $v_0=100$, $d_0=1$, $v_1=1$, $d_1=0.01$, $k_1=1$, $k_2=...=k_{N+1}=0.01$}.}\label{revplot3}
\end{figure}
We consider an arbitrarily large gene regulatory network which has been previously studied as a model of a circadian oscillator \cite{gonze2002robustness,thomas2013signatures}. The mechanism is as follows. A protein $P_1$ is translated by mRNA, $M$, which is itself transcripted by a gene in the on-state, $D_\text{on}$. Subsequently the protein $P_1$ generates $P_2$, and $P_2$ generates $P_3$, etc until a final protein $P_N$ is generated. The latter can bind to $D_\text{on}$ to deactivate it as $D_\text{off}$, which can reversibly unbind into $P_N$ and $D_\text{on}$. We seek the approximate distribution of the number of molecules of $D_\text{on}$ and $M$, which we expect to be accurate when the proteins are abundant. The full chemical system is described by the scheme:
\begin{align}\label{GenOsc}
D_\text{on}\xrightarrow{v_0}D_\text{on}+M,~M\xrightarrow{d_0} \emptyset,~M\xrightarrow{k_1}M+P_1,~P_1\xrightarrow{k_2}P_2\xrightarrow{k_3}...\xrightarrow{k_N}P_N\xrightarrow{k_{N+1}}\emptyset,~ D_\text{on}+P_N\xrightleftharpoons[v_1]{d_1}D_\text{off}.
\end{align}
\begin{figure}[h]
\centering
\includegraphics[scale=0.38]{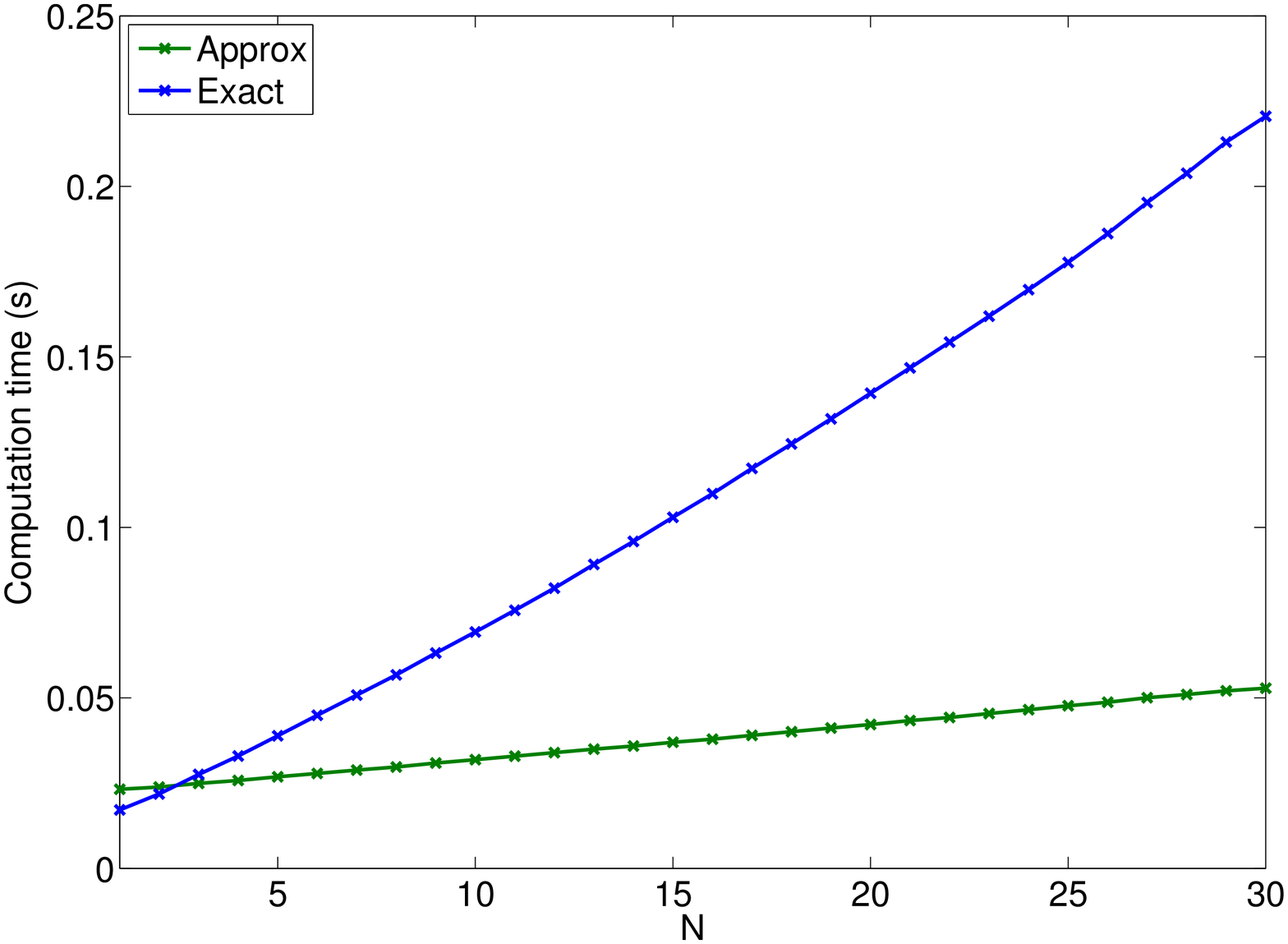}
\caption{{Computational time taken to compute an individual trajectory of length $10$ time units for the mRNA species in the genetic oscillator model using the Extrande algorithm for the reduced (approximate) chemical system (\ref{GenOscA}) and the SSA for the full (exact) chemical system (\ref{GenOsc}). The computational time is plotted as a function of the number of distinct protein species ($N$). Parameters are $\Omega=1$, $v_0=100$, $d_0=1$, $v_1=1$, $d_1=0.01$, $k_1=1$, $k_2=...=k_{N+1}=0.01$. Simulations were performed using MATLAB on a computer with a 3GHz Intel Core 2 Duo processor and 4GB RAM. Note that as the total number of protein species increases, stochastic simulations of the reduced system (approximate) becomes significantly more computationally efficient than the SSA for the full system (exact).} }\label{revplot4}
\end{figure}
According to our method, encapsulated by Eq. (\ref{redchemsysN}), the reduced chemical system takes the simpler form:
\begin{align}
\label{GenOscA}
D_\text{on}\xrightarrow{v_0}D_\text{on}+M,~M\xrightarrow{d_0} \emptyset,~ D_\text{on}\xrightleftharpoons[v_1]{d_1\phi_{P_N}(t)}D_\text{off}.
\end{align}
We note that the distribution of $D_\text{on}$ is independent of $M$, and is therefore simply Bernoulli$\left( \frac{v_1}{d_1\phi_{P_N}(t)+v_1}\right)$. The steady-state distribution of $M$ can be straightforwardly obtained using the method in Ref. \cite{grima2012steady} or else a fast implementation of the finite-state projection algorithm is equally effective \cite{munsky2006finite}. For a time-dependent description, however, we must use stochastic simulations to determine the accuracy of our method. In Fig. \eqref{revplot3} we plot the time development of the distribution of the number of mRNA molecules, $M$, for parameters such that the steady state is characterised by a fixed large abundance of proteins, in particular, $\frac{\phi_{P_i}}{\phi_M}=100,  \forall i$, which is a physically realistic ratio for some cells \cite{taniguchi2010quantifying}. Remarkably the approximate distribution provides an excellent match to the exact distribution for all times, reproducing even the transition from unimodality to bimodality and back to unimodality as a function of time.}

{In Fig. \eqref{revplot4} we plot the computational time taken to simulate an individual trajectory of length 10 time units with the SSA for the full system (\ref{GenOsc}) and Extrande for the approximate system (\ref{GenOscA}). For the approximate system, only 4 reactions must be simulated, while the full system has $N+5$ reactions. On the other hand, the approximate system requires the time-dependent solution of an $N+2$ dimensional system of REs. This trade-off implies that for $N=1,2$, the full system is slightly faster, but for any $N>3$, the approximate system is faster. 
}
\subsection{Exact Reductions}
We have already shown that our approximation is exact in the limit of infinite concentration of the abundant species, but as we now show, surprisingly, there is also a wide class of systems where the method is exact, regardless of abundance separation. 

\subsubsection{Systems in Detailed Balance}
We now show that for systems in detailed balance \cite{VKbook}, where the species we remove from the system (what we would previously call ``abundant" species) are not involved in a chemical conservation law, the approximation is exact. {Note that detailed balance is a property of some systems in steady-state conditions, and hence necessarily, the exactness mentioned does not apply to finite times, rather it applies only in the limit of infinitely long times.}

Consider a detailed balance system of $R$ reversible reactions, and $N$ chemical species, $X_1,...,X_N$. Let us denote reaction $j$ as:
\begin{align}\label{detailed}
s_{1j}X_1+...+s_{Nj}X_N \xrightleftharpoons[k_j']{k_j} r_{1j}X_1+...+r_{Nj}X_N,
\end{align}
where we allow a conservation law on the species $X_1,...,X_M$ of the form:
\begin{equation}
\sum_{i=1}^M\alpha_in_i=k,
\end{equation}
where $\alpha_i$ and $k$ are time-independent constants. Application of the method described by van Kampen in \cite{vk1976}, leads to an explicit expression for the steady-state solution of the CME of this chemical system which is given by a constrained multivariate Poisson distribution of the form:
\begin{equation}
P(\bold{n})=C \frac{\left(\Omega \phi_1\right)^{n_1}...\left(\Omega \phi_N\right)^{n_N}}{n_1!...n_N!} \delta \left( \alpha_1 n_1 + ... + \alpha_M n_M , k \right),
\end{equation}
where $\phi_i$ are the steady-state rate equation solutions, $\delta(...,...)$ is a Kronecker delta and $C$ is a normalisation constant. The marginal distribution is obtained by summing over $n_{M+1},...,n_N$. The exact marginal is therefore,
\begin{equation}\label{exactmargdb}
P^\star (\bold{n}')=\sum_{n_{M+1}=0}^\infty ... \sum_{n_N=0}^\infty  P(\bold{n})= C' \frac{\left(\Omega \phi_1\right)^{n_1}...\left(\Omega \phi_M\right)^{n_M}}{n_1!...n_M!} \delta \left( \alpha_1 n_1 + ... + \alpha_M n_M , k \right),
\end{equation}
where $C'$ is a normalisation constant.

Now the approximate reduction introduced in Section 2 is equivalent to approximating the chemical system (\ref{detailed}) by the reduced chemical system: 
\begin{align}
s_{1j}X_1+...+s_{Mj}X_M \xrightleftharpoons[k_j'\phi_{M+1}^{r_{M+1,j}}...\phi_{N}^{r_{N,j}} ]{k_j\phi_{M+1}^{s_{M+1,j}}...\phi_{N}^{s_{N,j}}}r_{1j}X_1+...+r_{Mj}X_M,
\end{align}
with the same conservation law as above. Application of the method in \cite{vk1976} to the reduced CME describing the above system, immediately leads to a steady-state solution which is exactly the same as Eq. \eqref{exactmargdb} since the RE solution of the reduced system is the same as that of the full system. Hence the approximation is exact for this class of chemical systems in detailed balance.

\subsubsection{Open Michaelis-Menten reaction with one enzyme molecule}
In the following example, we show that the approximation can be exact in steady-state conditions without taking any abundant limits, even if the system is not in detailed balance.

The open Michaelis-Menten reaction is given by:
\begin{align}
\label{OMM}
\emptyset \xrightarrow{k_\text{in}}S,~S+E\xrightleftharpoons[k_1']{k_1}C\xrightarrow{k_2}E+P,
\end{align}
where substrate molecules $S$ are input into the system, they reversibly bind with enzyme $E$ to form a complex $C$ which in turn irreversibly decays into the original enzyme $E$ and product molecules $P$. 

We will consider the case with the conservation law $n_E+n_C=1$, that is where there is just one enzyme molecule in the compartment. The CME describing the above reaction system is:
\begin{align}
\dot{P}(n_S,n_E)&=k_\text{in}\Omega (P(n_S-1,n_E)-P(n_S,n_E) ) +\frac{k_1}{\Omega} ( (n_S+1)(n_E+1)P(n_S+1,n_E+1)-n_Sn_EP(n_S,n_E))\nonumber\\
&+k_1' ((1 -n_E+1)P(n_S-1,n_E-1)-(1-n_E)P(n_S,n_E))\nonumber\\
&+k_2((1-n_E+1)P(n_S,n_E-1)-(1-n_E)P(n_S,n_E)).
\end{align}
This equation has been solved exactly in steady-state conditions in Appendix G of Schnoerr et al. \cite{schnoerr2014}. In particular therein it was shown that the average enzyme molecule number in steady-state condition is given by $\langle n_E \rangle=1-\frac{k_\text{in}\Omega}{k_2}$. This together with the fact that a single enzyme molecule, at any given time, can be in only one of two states, implies that the steady-state marginal distribution of enzyme number fluctuations is:
\begin{align}
\label{Bres}
P^{\star}(n_E) \sim \text{Bernoulli} \left( 1-\frac{k_\text{in}\Omega}{k_2} \right).
\end{align}

Next we show that our reduction gives exactly the same distribution, regardless of the abundance of substrate. The reduced CME describing enzyme fluctuations is given by:
\begin{align}
\dot{\tilde{P}}(n_E)&=k_1 \phi_S \left( (n_E+1)\tilde{P}(n_E+1)- n \tilde{P}(n_E) \right)+(k_1'+k_2) \left( (2-n_E)\tilde{P}(n_E-1)-(1-n)\tilde{P}(n_E) \right).
\end{align}
In steady state, setting $n_E=0$ gives us:
\begin{align}
k_1  \phi_S  \tilde{P}(1)-(k_1'+k_2 ) \tilde{P}(0)=0.
\end{align} 
Therefore, with the condition $\tilde{P}(0)+\tilde{P}(1)=1$, we find that,
\begin{align}\label{eq59}
\langle n_E \rangle=\tilde{P}(1)=\frac{k_1'+k_2}{k_1\phi_S+k_1'+k_2}.
\end{align}
The REs for this system are:
\begin{align}
\dot{\phi}_S&=k_\text{in}+k_1'\Big(\frac{1}{\Omega}-\phi_E\Big)-k_1\phi_E \phi_S,\nonumber\\
\dot{\phi}_E&=-k_1\phi_E\phi_S+(k_1'+k_2)\Big(\frac{1}{\Omega}-\phi_E\Big),
\end{align}
which possess a steady state solution:
\begin{align}\label{eq61}
\phi_S=\frac{k_\text{in}}{k_1} \left( \frac{k_1'+k_2}{k_2-k_\text{in}} \right).
\end{align}
Substituting Eq. \eqref{eq61} into Eq. \eqref{eq59}, we find,
\begin{align}
\langle n_E \rangle = \tilde{P}(1)=1-\frac{k_\text{in}\Omega}{ k_2} = \Omega \phi_E.
\end{align}
As by arguments before, the steady-state distribution is Bernoulli and hence it follows that:
\begin{align}
\tilde{P}(n_E) \sim \text{Bernoulli} \left( 1-\frac{k_\text{in}\Omega}{k_2} \right),
\end{align}
which is equal to the exact solution Eq. (\ref{Bres}).

By similar arguments, it can be easily deduced that the marginal distribution of any species which exists in two states and for which the average number of molecules predicted by the REs is the same as the CME, is exactly predicted by the reduced CME. The second criterion on average molecule numbers is bound to generally be the limiting one since it is typically not the case that the REs exactly agree with the mean concentrations calculated from the CME (see for example Ref. \cite{grima2010}). For example for the genetic feedback loop (\ref{genfeedeq}) the marginal distributions of the gene in the on or off state cannot be exactly predicted by the reduced CME because as shown in Ref. \cite{grima2012steady}, the average number of genes in each state (equivalently the fraction of time spent in each state) predicted by the CME does not equal that of the REs. 

\section{Estimating the approximation error of the hybrid model}

As we have shown for most systems, the reduction is exact only in the limit of infinite concentrations of certain species, and the reduction is therefore an approximation if concentrations are finitely large. 

{We now investigate the use of the Linear Noise Approximation (LNA) to obtain an estimate of the error made by the use of the reduced CME. By comparing this estimate with that obtained from stochastic simulations of both the full and approximate systems, we demonstrate that the LNA's estimate is accurate for a wide range of parameters and systems. Since the LNA is obtained by solving a system of coupled ordinary differential equations, our results suggest the use of the LNA as a computationally efficient means of estimating the error which bypasses lengthy stochastic simulations using stochastic simulations of the full and reduced CMEs.}

The LNA is an approximation which assumes the fluctuations in each chemical species are normally distributed. More precisely, it is the leading order approximation of the system-size expansion of the CME \cite{VKbook} in the limit of large volumes. The general formulation of the LNA is as follows (see for example \cite{elf2003fast} for more details).  

Consider a system of $N$ chemical species, with $R$ reactions, where the $j^\text{th}$ reaction is:
\begin{equation}
\label{grs}
s_{1j}X_1+...+s_{Nj}X_N \xrightarrow{k_j} r_{1j}X_1+...+r_{Nj}X_N.
\end{equation}
The REs for the system are then given by:
\begin{equation}\label{68}
\dot{\vec{\phi}}=S \vec{f},
\end{equation}
where we remind the reader that $S$ is the stochiometric matrix with elements $S_{ij} = r_{ij}-s_{ij}$ and $\vec{f}$ is the macroscopic propensity vector $\vec{f}$ defined as:
\begin{equation}\label{67}
f_j=k_j\phi_1^{s_{1j}}...\phi_N^{s_{Nj}}.
\end{equation}

The Jacobian matrix $J$ is the derivative of Eq. \eqref{68} with respect to $\vec{\phi}$:
\begin{equation}\label{69}
J=S \frac{\partial \vec{f}}{\partial \vec{\phi}}.
\end{equation}
The diffusion matrix $D$ is given by the matrix product:
\begin{equation}
\label{69a}
D=S ~\text{diag}(\vec{f} ) ~S^T.
\end{equation}
The time-evolution of the second moments of the fluctuations is then approximately given by the Lyapunov differential equation:
\begin{equation}\label{70}
\frac{dC}{dt}=JC+CJ^T+D,
\end{equation}
where $\Omega C_{ij}$ is the LNA estimate for the CME's prediction of the covariance in the number fluctuations of species $X_i$ and $X_j$. 

Now the proposed reduction approximates reaction scheme (\ref{grs}) by:
 \begin{align}
s_{1j}X_1+...+s_{Mj}X_M \xrightarrow{k_j\phi_{M+1}^{s_{M+1,j}}...\phi_{N}^{s_{Nj}}}r_{1j}X_1+...+r_{Mj}X_M, \quad j = 1,..., R,
\end{align}
when species $X_{M+1},...,X_N$ are the abundant species. Note that for the reduced system with $M$ species, the abundant concentrations no longer function as concentrations, but instead as parameters like the reaction rates $k_j$. The REs remain unchanged, however. The Jacobian and diffusion matrices for this system, $\tilde{J}$ and $\tilde{D}$ , are hence the upper left $M \times M$ blocks of $J$ and $D$ previously defined in Eqs. (\ref{69}) and (\ref{69a}), respectively. Thus the LNA leads to an estimate of the reduced CME's prediction of the covariance of fluctuations, $\Omega \tilde{C}$, which is the solution of the Lyapunov equation:
\begin{equation}\label{70a}
\frac{d\tilde{C}}{dt}=\tilde{J}\tilde{C}+\tilde{C}\tilde{J}^T+\tilde{D}.
\end{equation}

Hence it follows that the LNA's estimate of the absolute relative difference in the variance predictions of the full and reduced CME's for species $i$ is given by:
\begin{align}\label{Relerr}
R_i=\frac{\vert C_{ii} - \tilde{C}_{ii} \vert }{C_{ii}}.
\end{align}
Of course one can also calculate this quantity as a function of time, by solving the Lyapunov equations of the full and reduced CMEs numerically; however in what follows we shall assume steady-state conditions to simplify the presentation. 

Though its generally impossible to obtain a closed-form simple analytical solution to the LNA equations, one can show that the error $R_i$ is approximately proportional to the inverse of the ratio of the abundant to non-abundant species concentrations in the abundant limit. The proof is as follows. 

Referring to Table 2, if species $X_N$ is abundant, then the abundant limit consists of the rate constant scalings: $k_{L+N} \propto \frac{1}{x^2}$, $k_N \propto \frac{1}{x}$, $k_j \propto\frac{1}{x}$ for $j$ such that $a(j)$ or $b(j)$ equal $N$ ($j$ denotes a bimolecular reaction which involves $X_N$ and another species) and the steady-state concentration scalings: $\phi_i = c_i$ for $i \ne N$, where $c_i$ are constants independent of $x$ and $\phi_N \propto x$. It is easy to verify using this limit and the REs given by Eq. (\ref{eqreN}) that the Jacobian matrix can be written as $J = J_0 + y J_1$, where $y = 1/x$ and $J_i$ are matrices to be determined from the REs. In particular $J_0$ is the Jacobian of the REs with the terms describing the removal of the abundant species set to zero. The same scaling form for the Jacobian is obtained for any number of abundant species.

On the other hand, the diffusion matrix $D$ is unchanged under the abundance limit. This is since by Eq. (\ref{69a}), the elements of the $D$ are linear functions of the macroscopic rate functions $f_j$ (see Eq. (\ref{67})) which are unchanged by the abundance limit since each limit of $k_j$ tending to zero will be counterbalanced by the opposite limit of a concentration of an abundant species tending to infinity. 

Hence in the abundance limit, the Lyapunov Eq. (\ref{70}) can be written as:
\begin{equation}
\frac{dC}{dt}=(J_0 + J_1 y)C+C(J_0^T + J_1^T y)+D.
\end{equation}
The form of this equation suggests a solution of the type $C = C^{(0)} + C^{(1)} y + C^{(2)} y^2 + O(y^3)$. Indeed plugging this ansatz in the above equation, one transforms it into a coupled set of equations for the matrices $C^{(i)}$ which can be solved iteratively, i.e., the equation for $C^{(i)}$ depends on $C^{(j)}$ where $j < i$ except for $C^{(0)}$ which is a function of $J_0$ and $D$ only. Now the abundance limit is the limit $y = 1/x \rightarrow 0$ and hence the relative error in the variance can be written as:
 \begin{align}\label{Relerr1}
R_i=\frac{\vert C_{ii} - C_{ii}^{(0)} \vert }{C_{ii}} = \frac{\vert C_{ii}^{(1)} \vert}{C_{ii}^{(0)}} y + O(y^2).
\end{align}
Now in the abundance limit, the ratios of abundant to non-abundant concentrations are proportional to $x = 1/y$ and hence it follows that in this limit, the relative error $R_i$ is proportional to the inverse of these ratios. 

Next we demonstrate the accuracy of the LNA's estimate of the relative error in the predictions of the reduced CME, i.e., Eq. (\ref{Relerr}). This is done by comparison of the LNA estimate with the relative error directly computed from the SSA of the full CME and the steady-state analytical solution of the reduced CME for three examples of biochemical relevance. 

\subsection{Open Michaelis-Menten reaction with multiple enzyme molecules}

We consider the open Michaelis Menten system (\ref{OMM}) with multiple number of enzyme molecules, i.e.,  $(n_E + n_C) = E_T$, where $E_T$ is the total number of enzyme molecules. We consider the case where the substrate is much more abundant than the enzyme. Computing the LNA of the full CME and of the reduced CME for the non-abundant enzyme species, we find that the relative error in the variance of the enzyme number fluctuations, as given by Eq. \eqref{Relerr}, is:
\begin{align}
R=\frac{(1-a)\frac{E_T}{\Omega}+K_M}{(1-a)^2\frac{E_T}{\Omega}+K_M} - 1,
\end{align}
where $a=\frac{k_\text{in}\Omega}{k_2 E_T}$ and $K_M=\frac{k_1'+k_2}{k_1}$ is the Michaelis-Menten constant.

The steady-state substrate concentration solution of the REs for this system is:
\begin{align}
\phi_S=\frac{K_M a}{(1-a)}.
\end{align}
We define $L=\frac{\Omega\phi_S}{E_T}$ as a measure of the relative abundance of substrate $S$. It then follows that $R$ can be written as:
\begin{align}\label{Req}
R=\frac{a^2}{a(1-a)+L}.
\end{align}
The condition $0<a<1$ is a requirement for the existence of a steady state, and $R$ is a monotonically increasing function of $a$, so the maximum possible value of $R$ is at $a=1$, in other words,
\begin{align}
R_\text{max}=\frac{1}{L}=\frac{E_T}{\Omega\phi_S}.
\end{align}
That is, if the substrate concentration is ten times the total enzyme concentration, then the percentage relative error in the reduced CME's estimate of the variance of enzyme number fluctuations will be less than ten percent.

\begin{figure}
\includegraphics[scale=0.41]{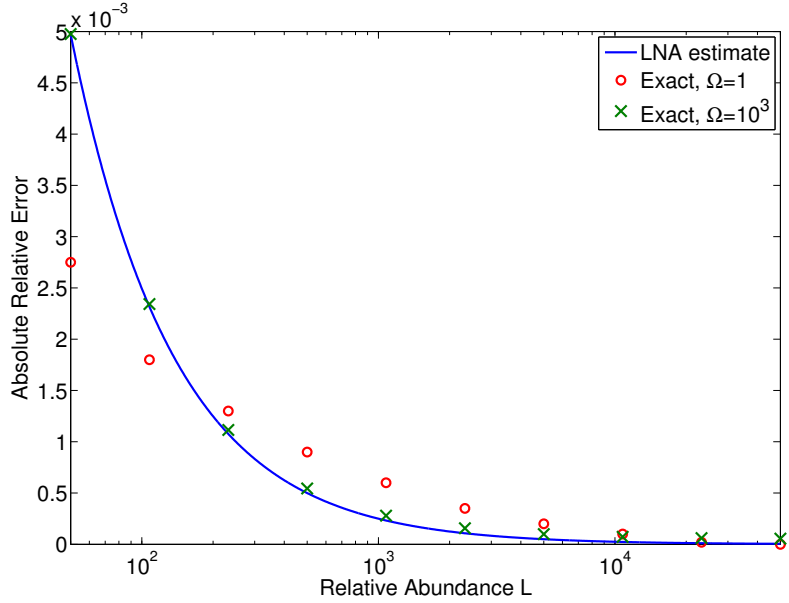}
\caption{Relative error in the reduced CME's prediction of the variance of fluctuations of non-abundant species for the open Michaelis-Menten reaction \eqref{OMM} in steady-state conditions. The true relative error in the variance of non-abundant enzyme number fluctuations at $\Omega=1$ (red circles) and $\Omega=10^3$ (green crosses) computed using a time-average of SSA (for the full CME) and the analytical steady-state distribution solution of the reduced CME, and compared with the LNA estimate given by Eq. \eqref{Req} (blue line). For this example, the LNA gives a reasonable estimate of the true error at $\Omega=1$ and an excellent estimate at $\Omega=10^3$. $L$ is the relative abundance defined as the substrate concentration divided by the total enzyme concentration (according to the REs). Parameter values are $k_\text{in}=10,~k_1=\frac{1}{x},~k_1'=10^2,~k_2=1,~E_T=20,~10<x<10^4$.}\label{fig4}
\end{figure}

The reduced CME can in this case be exactly solved in steady-state conditions and one obtains a binomial distribution with parameters $E_T$ and $1-a$ describing the fluctuations in enzyme molecule numbers; indeed for the case $E_T = 1$, the binomial distribution reduces to the Bernoulli distribution found earlier for the open Michaelis Menten system with one enzyme molecule (see Eq. (\ref{Bres})). In Fig. \eqref{fig4} we use the variance calculated from this solution together with the variance calculated from time-averages of SSA (for the full CME) to compute the true error in the reduced CME's variance of enzyme number fluctuations for the open Michaelis-Menten system. This is done for two different volumes, $\Omega=1$ and $\Omega=10^3$. The true error is also compared in the same figure with the LNA estimate given by Eq. \eqref{Req}. The relative concentrations of substrate and enzyme are controlled by setting the rate constant $k_1$ proportional to  $1/x$ and varying $x$ (in accordance with the abundance limits discussed earlier; see caption of Fig. 4 for details). The LNA estimates are reasonably good for both volumes but practically indistinguishable from the true error for the larger volume of $\Omega=10^3$. This is to be expected since the LNA becomes exact in the limit of large volumes. 

\subsection{Open Homodimerisation reaction}

Next we use the LNA to estimate the errors in the reduced CME description for the homodimerisation example (\ref{homodimeq}).  We consider the case in which species $X_1$ is abundant compared to species $X_2$. Choosing the scalings $k_2 = c_1 / x^2$ and $k_3 = c_2 / x$ (where $c_i$ are proportionality constants), it follows by the considerations of Section 2.2 that the steady-state concentration of $X_1$ is proportional to $x$ while that of $X_2$ is a constant; hence by varying $x$ we have a convenient way to control the ratio of the two concentrations. In particular in steady-state, the solution of the REs is given by:
\begin{equation}
\phi_1 = \frac{\sqrt{c_2^2 + 8 c_1 k_1} - c_2}{4 c_1} x, \phi_2 = \frac{c_2^2 + 4 c_1 k_1 - c_2 \sqrt{c_2^2 + 8 c_1 k_1}}{8 c_1 k_4}.
\end{equation}
The LNA relative error in the variance of the fluctuations of the non-abundant species $X_2$ as given by Eq. (\ref{Relerr}) is:
\begin{equation}
\label{errorOHR}
R =\frac{\lambda_0}{\lambda_1 + x \lambda_2},
\end{equation}
where $\lambda = (c_2 + \sqrt{c_2^2 + 8 c_1 k_1}) k_4 / 2 c_1 k_1$, $\lambda_0 = 4 c_1^2 \phi_2 \lambda^2 (k_1 - \phi_2 \lambda (c_2 + 4 c_1 \phi_2 \lambda))$, $\lambda_1 = c_2^2 k_4 + 8 c_1 c_2 \phi_2 k_4 \lambda + c_1 c_2^2 \phi_2 \lambda^2 + 4 c_1^2 \phi_2 (k_1 + 4 \phi_2 k_4) \lambda^2 + 4 c_1^2 c_2 \phi_2^2 \lambda^3$ are constants. Note that the ratio of the abundant to the non-abundant species concentrations is proportional to $x$. Hence by Eq. (\ref{errorOHR}) it follows that the relative error has a maximum equal to $\lambda_0 / \lambda_1$ and decreases monotonically as $X_1$ becomes more abundant relative to $X_2$. Next we test the accuracy of the LNA estimate. 

\begin{figure}[h]
\includegraphics[scale=0.41]{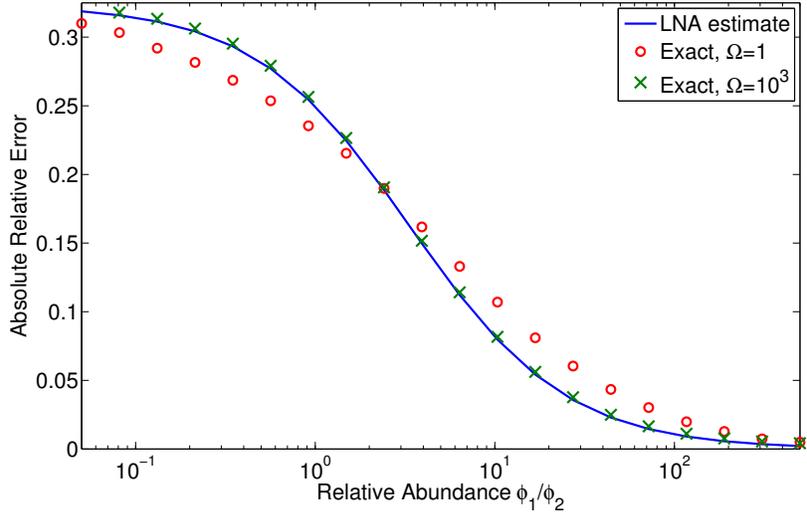}
\caption{Relative error in the reduced CME's estimate of the variance of fluctuations of non-abundant species for the open homodimerisation reaction \eqref{homodimeq} in steady-state conditions. The true relative error in the variance of the fluctuations of the non-abundant species $X_2$ at $\Omega=1$ (red circles) and $\Omega=10^3$ (green crosses) are computed using the time-averages of SSA for the full CME and the analytical steady-state distribution of the reduced CME, and compared with the LNA estimate given by Eq. \eqref{errorOHR} (blue line). For this example, the LNA gives a good estimate of the true error for $\Omega=1$ and an excellent estimate for $\Omega=10^3$. Parameter values are $k_1=10^3$, $k_4=10$, $k_2=\frac{100}{x^2}$, $k_3=\frac{10}{x}$, where $1<x<10^4$.}\label{fig5}
\end{figure}

The reduced CME for this system can be exactly solved in steady-state conditions and one obtains a Poisson distribution for the fluctuations in the number of molecules of $X_2$ with parameter $\Omega k_2 \phi_1^2/k_4$. In Fig. \ref{fig5} we use the variance calculated from this solution together with the variance calculated from time-averages of SSA (for the full CME) to compute the true error in the reduced CME's variance of number of $X_2$ fluctuations. This is done for two different volumes, $\Omega=1$ and $\Omega=10^3$. The true error is also compared in the same figure with the LNA estimate given by Eq. \eqref{errorOHR}. As for the previous example, the LNA accuracy is good across a wide range of volumes and becomes particularly accurate in the limit of large volumes. It is also noteworthy that the LNA estimate of the relative error is good over a wide range of relative abundances; in particular it even provides an accurate value (about 0.3) for the maximum relative error which occurs in the limit of small relative abundance of $X_1$ compared to $X_2$.

\subsection{Genetic Feedback Loop}

Here we use the LNA to estimate the error in the reduced CME description for the genetic feedback loop (\ref{genfeedeq}), where the gene concentration is fixed to $1/\Omega$, i.e, a single gene. We consider the case where the protein $P$ is much more abundant than the gene. Choosing the scalings $d_0 = c_0 / x$ and $d_1 = c_1 / x$ (where $c_i$ are proportionality constants), it follows by the considerations of Section 2.2 that the steady-state concentration of $X_2$ (the protein) is proportional to $x$ while that of $X_1$ (the gene) is a constant; hence by varying $x$ we have a convenient way to control the ratio of the two concentrations. In particular in steady-state, the solution of the REs is given by:
\begin{align}
\phi_1 &= \frac{2 c_1 v_0 + c_0 \Omega v_1 - \sqrt{(c_0 \Omega v_1) (4 c_1 v_0 + c_0 \Omega v_1)}}{2 c_1 \Omega v_0}, \\ \phi_2 &= (2 c_1)^{-1} x \biggl(\frac{ \sqrt{v_1 (4 c_1 v_0 + c_0 \Omega v_1)}}{\sqrt{c_0 \Omega}} - v_1 \biggr).
\end{align}
The LNA relative error in the variance of the fluctuations of the non-abundant gene as given by Eq. (\ref{Relerr}) is:
\begin{equation}
\label{errorGC}
R =\frac{\lambda_0}{\lambda_1 + x \lambda_2},
\end{equation}
where $\lambda = \phi_2 / \phi_1 x$, $\lambda_0 =-c_1 (-1 + \phi_1 \Omega) (c_1 (-1 + \phi_1 \Omega) (-1 + 2 \phi_1 \Omega) v_0 v_1 - c_0 \phi_1 \Omega^2 v_1 (v_0 + v_1) + c_0 c_1 \phi_1 \Omega ((-1 + \phi_1 \Omega) v_0 - \phi_1 \Omega v_1) \lambda)$, $\lambda_1 = (v_1 + c_1 \phi_1 \lambda) (c_0^2 \phi_1 \Omega^3 v_1 + c_0^2 c_1 \phi_1 \Omega^2 \lambda - c_0^2 c_1 f1^2 \Omega^3 \lambda + c_1^2 (-1 + \phi_1 \Omega) ((-1 + \phi_1 \Omega) v_0 - c_0 \phi_1 \Omega \lambda) - c_1^2 \phi_1 \Omega (-1 + \phi_1 \Omega) ((-1 + \phi_1 \Omega) v_0 - c_0 \phi_1 \Omega \lambda))$ and $\lambda_2 = (v_1 + c_1 \phi_1 \lambda) (c_1 \phi_1 \Omega^2 (1 - \phi_1 \Omega) v_0 v_1 + c_0 \phi_1 \Omega^3 v_1^2 + c_0 c_1 \phi_1 \Omega^2 v_1 \lambda + c_1^2 \phi_1 \Omega (-1 + \phi_1 \Omega) \lambda ((-1 + \phi_1 \Omega) v_0 - c_0 \phi_1 \Omega \lambda))$ are constants. Note that the ratio of the abundant to the non-abundant species concentrations is proportional to $x$. Hence by Eq. (\ref{errorGC}) it follows that the relative error has a maximum equal to $\lambda_0 / \lambda_1$ and decreases monotonically as $X_2$ becomes more abundant relative to $X_1$. Note that the form of the LNA estimate of the error in this example is the same as that in the previous example. 

\begin{figure}[h]
\includegraphics[scale=0.41]{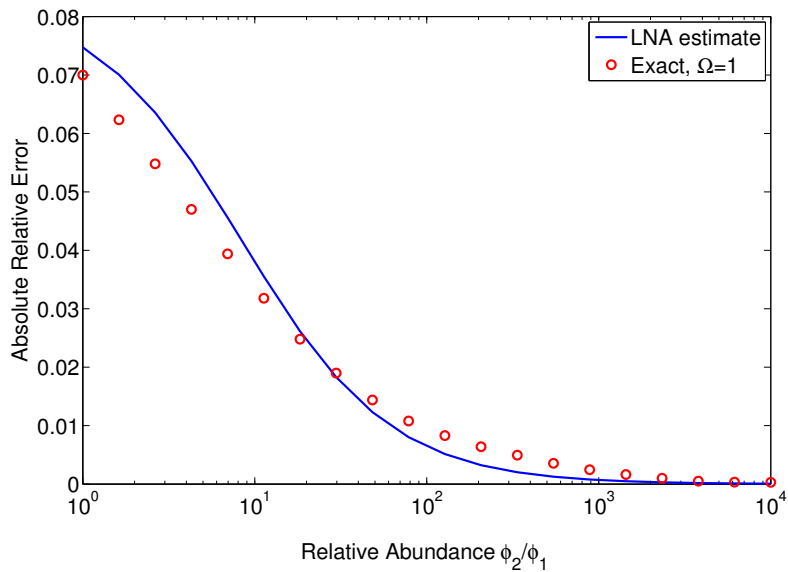}
\caption{Relative error in the reduced CME prediction of the variance of gene fluctuations for the genetic feedback loop \eqref{genfeedeq} in steady-state conditions. The true relative error at $\Omega=1$ (red circles) is computed using a time-average of SSA for the full CME and the analytical steady-state solution of the reduced CME, and compared with the LNA estimate given by Eq. \eqref{errorGC} (blue line). Parameter values are $v_0=1$, $v_1=0.1$, $d_0=\frac{1}{x}$, $d_1=\frac{0.01}{x}$, where $1<x<10^4$. The relative abundance of protein to gene concentrations is $\phi_2 / \phi_1$ (according to the REs) }\label{fig6}
\end{figure}

In Fig. \ref{fig6} we plot the true error in the variance of the fluctuations of the non-abundant gene computed using time-averages of SSA (for the full CME) and the analytical solution of the reduced CME in steady-state conditions (a Bernoulli distribution with parameter given by the steady-state solution of Eq. (\ref{thetaeq})) at $\Omega=1$. This is compared with the LNA estimate Eq. \eqref{errorGC} which is found to be particularly accurate, as found previously for the enzyme and dimerisation examples. However unlike the previous examples, for the gene system, in the limit of large $\Omega$, the LNA's estimate does not become more accurate. The reason is that the LNA is accurate in the deterministic limit in which all species molecule numbers increase to infinity at constant concentration whereas in this example the gene molecule number is fixed to one and only the protein molecule number is taken to infinity. 

{
\subsection{Genetic oscillator}
Finally, we use the LNA to estimate the error in the reduced CME description for the genetic oscillator (\ref{GenOsc}). We consider the case where the proteins $P_i$  are more abundant than the mRNA, while the gene is restricted to a maximum concentration of $\frac{1}{\Omega}$, i.e, a single molecule. Choosing the scalings $d_1 = c_1 / x$ and $k_2 = ...=k_{N+1}=c_2 / x$ (where $c_i$ are proportionality constants), it follows by the considerations of Section 2.2 that the steady-state concentrations of $P_i$ (the proteins) is proportional to $x$ while that of $D_\text{on}$ (the gene) and $M$ (the mRNA) are constant; hence by varying $x$ we have a convenient way to control the ratio of the two concentrations. In particular in steady-state, the solution of the REs is given by:
\begin{align}
\phi_{D_{on}} &=\frac{\sqrt{v_1^2+4v_0c_1k_1v_1/(d_0c_2\Omega)-v_1}}{2v_0c_1k_1/(d_0c_2)}, \\ \phi_M &= \frac{v_0}{d_0}\phi_{D_{on}},\\
\phi_{P_i}&=\frac{k_1x}{c_2}\phi_M.
\end{align}
Given the arbitrarily large number of species, there is no compact analytic expression for the LNA relative error in the variance of the fluctuations of the non-abundant mRNA as given by Eq. (\ref{Relerr}), however the error can be calculated by numerical solution of the REs and the Lyapunov equations of the full and reduced systems.}

\begin{figure}[h]
\includegraphics[scale=0.41]{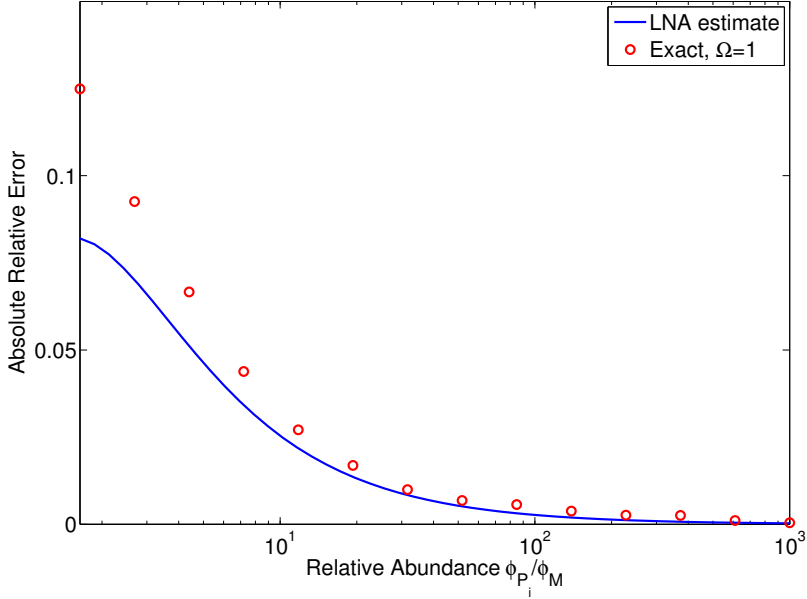}
\caption{{Relative error in the reduced CME prediction of the variance of mRNA fluctuations for the genetic oscillator \eqref{GenOsc} in steady-state conditions. The true relative error at $\Omega=1$ (red circles) is computed using a time-average of the SSA for the full CME and the steady-state solution of the reduced CME computed with the finite-state projection algorithm, and compared with the LNA estimate (blue line). Parameter values are $N=10$, $v_0=100$, $v_1=1$, $d_0=1$, $d_1=\frac{1}{x}$, $k_1=1$, $k_2=...=k_{N+1}=\frac{1}{x}$ where $1<x<10^3$. The relative abundance of protein to mRNA concentrations is $\phi_{P_i} / \phi_M$ (according to the REs) which is the same for all $i=1,...,N$.}}\label{fig8}
\end{figure}

{In Fig. \ref{fig8} we plot the true error in the variance of the fluctuations of the non-abundant mRNA computed using time-averages of SSA (for the full CME) and the solution of the reduced CME in steady-state conditions (computed with the finite-state projection algorithm) at $\Omega=1$. This is compared with the LNA estimate which is found to be accurate for physically realistic abundances (ratio of protein to mRNA concentrations are commonly larger than a hundred in bacteria; see for example \cite{taniguchi2010quantifying}). However unlike some of the previous examples, for the gene system, in the limit of large $\Omega$, the LNA's estimate does not become more accurate. The reason is that the LNA is accurate in the deterministic limit in which all species molecule numbers increase to infinity at constant concentration whereas in this example the gene molecule number is fixed to one and only the protein molecule number is taken to infinity.}

 \section{Summary and Discussion}
Summarising, in this paper we have introduced a novel reduced stochastic description of chemical systems in which some species are abundant. The key intuitive idea is to replace the conditional expectation of the number of molecules of abundant species in the propensities of the exact marginal CME by the solutions of the deterministic REs, and hence obtain a reduced CME for the non-abundant species only. Therefore our method is a hybrid approach. We note that our method is different and simpler than that presented in \cite{hellander2007,Jahnke2011,Jahnke2012} since the latter postulate a more complicated approach than REs to model the abundant species. This relative simplicity indeed leads to three major strengths of our approach over existing approaches: (i) it is easier to implement and computationally more efficient than present approaches; (ii) our reduced CME can be explicitly solved for a number of biochemically relevant examples; (iii) simple rational expressions can be derived which estimate the errors inherent in the hybrid approximation relative to the fully stochastic description. Curiously we also found that the reduced CME at the heart of our hybrid method is exact for some chemical systems, i.e., without requiring the necessity of abundance separation or without restricting the system to purely monomolecular systems (as was found to be the case in \cite{Jahnke2011} to ensure exactness for the Hellander and Lotstedt model). The major disadvantage of our approach is that its unlikely that it will be able to capture as many features of the fully stochastic model as the more sophisticated approaches mentioned above.

The present work also suggests some new avenues of research. The first is finding exact error bounds for the reduced CME, which could provide useful reassurance for mathematical modellers using this method. A second interesting direction would be to develop a more refined reduction of the CME by replacing the conditional expectation of the number of molecules of abundant species in the propensities of the exact marginal CME by the solutions of effective mesoscopic rate equations (EMREs) \cite{grima2010} instead of REs. EMREs have been demonstrated to be more accurate than REs in the sense that the difference between their mean concentration solution and that of the CME is considerably smaller than the difference between the RE solution and that of the CME \cite{Ramaswamy2012}. {Hence a CME reduction based on EMREs is highly likely to be more accurate than the one developed in this paper, particularly for cases where the compartment volume is small such that even though there is a large ratio of abundant to non-abundant concentrations, the number of molecules of abundant species is small}. Finally another interesting area for future work is the relationship between time scale separation and abundance separation. It is clear that latter does not typically imply the type of time scale separation typically used to obtain reduced CMEs (see for example \cite{Cao2005}) because it does not lead to a partitioning of reactions into fast and slow types; this is since within our abundance separation method, each limit of a rate constant tending to zero is counterbalanced by the opposite limit of a concentration of an abundant species tending to infinity. Yet it is not difficult to show that our abundance separation limit does lead to a separation of the eigenvalues of the Jacobian and hence point to timescale separation of concentration transients on the deterministic level. Thus a deeper investigation into the relationship between abundance separation and time scale separation could improve understanding of both types of separation and as well lead to a clearer picture regarding when CMEs can be effectively reduced.

\end{document}